# Integrated yoctosecond-precision timing detector

**Jie Yang,[1†] Tong Wang,[1†] Yulin Shen,[1] Dehui Pan,[1] Zhichao Chen,[2] Ming-Yang Zheng,[3*] Bin Wang,[3] Shaobo Fang,[4] Yi Zhang,[1] Ke Zhang,[1] Jiahui Yao,[1] Minghua Chen,[5*] Guorong Wu,[6*] Xueming Yang[2,6], and Ming Xin[1*]**

[1]*School of Electrical and Information Engineering, Tianjin University, Tianjin, 300072, China*

[2]*Dalian Institute of Chemical Physics, Dalian, 116023, China*

[3]*Jinan Institute of Quantum Technology, Jinan 250102, China*

[4]*Institute of Physics, Chinese Academy of Sciences, Beijing, 100190, China*

[5]*Department of Electronic Engineering, Tsinghua University, Beijing, 100084, China*

[6]*Institute of Advanced Light Source Facilities, Shenzhen 518107, China*

[†]These authors contributed equally to this work.

[*]Email: zhengmingyang@jiqt.org, chenmh@tsinghua.edu.cn, wugr@mail.iasf.ac.cn, xinm@tju.edu.cn

## ABSTRACT

Precise timing detection is essential for exploring ultrafast phenomena in fields ranging from free-electron lasers to ultra-high-power laser facilities. However, achieving simultaneous high resolution and large dynamic range remains a fundamental challenge, and state-of-the-art systems are often constrained by their physical size, power requirements, and limited scalability. Here we introduce an integrated dual electro-optic sub-cycle timing detector (DEST) that overcomes these limitations. In a proof-of-principle measurement, the device resolves timing jitter as small as 11 yoctoseconds (ys, $10^{-24}$ s) at 1 MHz—equivalent to the transit time of light across two protons—while maintaining an unambiguous measurement range of 6.15 ps and a dynamic range exceeding 155 dB. The core detection unit is miniaturized to chip-scale dimensions of 18 mm × 2 mm × 1 mm on a thin-film lithium niobate platform, ensuring inherent stability and immunity to environmental disturbances. Moreover, the architecture naturally lends itself to massive parallelization through array integration, with the potential to push timing precision to the sub-10 rontosecond (rs, $10^{-27}$ s) level within a 1 m² footprint. This combination of extreme sensitivity, wide dynamic range, compact size, and scalability opens new avenues for detecting previously inaccessible weak signals, including those from gravitational waves, quantum vacuum fluctuations, and beyond.

## INTRODUCTION

Timing refers to the measurement of time intervals between events or processes [1]. In experimental physics, engineering, and a wide range of technological domains, precise timing is essential for understanding and controlling phenomena that unfold over extremely short timescales [2]. To meet this need, timing detectors have been developed to record such intervals with high accuracy. Current timing detectors have reached remarkable levels of performance, with many systems now capable of operating at femtosecond and attosecond precision [3-10]. These advances have enabled critical applications in areas such as X-ray free-electron lasers [11-17], ultra-high-power laser facilities [18, 19], multi-telescope arrays [20], electron diffraction [21],waveform synthesis [22], soliton characterization [23], sub-nanometer ranging [24] and beyond [25].

Despite these advances, further progress in timing detection faces three fundamental challenges. First, achieving simultaneously sub-attosecond resolution and picosecond-scale measurement range remains extremely difficult for most existing detectors. Second, state-of-the-art systems such as those used in gravitational-wave observatories [26] rely on large-scale infrastructure; upgrading their sensitivity requires proportionally higher laser power and physical expansion of the facility, both of which face severe engineering and cost constraints. Third, the pathway to even higher precision—at the yoctosecond (ys, $10^{-24}$ s) level and beyond—has been hindered by both physical and practical constraints.

Here we introduce an integrated dual electro-optic sub-cycle timing detector (DEST) that overcomes these limitations. In a proof-of-principle measurement, by shifting the timing information to a radio-frequency (RF) sideband and using differential detection to reject common-mode noise, our device resolves timing jitter as small as 11 ys at 1 MHz—equivalent to the transit time of a photon across two protons—while achieving an average noise floor of 2.89 ys in 120 s. By uniquely identifying optical sub-cycles through a multidimensional voltage encoding scheme, the detector simultaneously offers an unambiguous measurement range of 6 ps and a dynamic range exceeding 155 dB. The core detection unit is miniaturized to chip-scale dimensions of just 18 mm × 2 mm × 1 mm, enabled by a thin-film lithium niobate platform. This compact integration not only ensures inherent stability and immunity to environmental disturbances but also makes the detector suitable for portable systems and harsh space environments.

In contrast to conventional large-scale facilities, where further improvements would require more powerful lasers and larger physical infrastructure, our approach leverages wafer-scale manufacturing and array integration—performance can be improved simply by adding more chips, rather than by scaling up a single device. Based on our calculations, this scalability could push timing precision to sub-10 rontoseconds (rs, $10^{-27}$ s) within a 1 $m^2$ footprint, opening up opportunities in particle physics [27, 28], quark-gluon plasma dynamics [29], space-based gravitational wave detection [30], quantum vacuum fluctuation [31] and diffraction [32], petahertz electronics [33], and potentially enabling the discovery of entirely new signals that remain beyond the reach of current technologies.

## THEORETICAL ANALYSIS

The principle of DEST is depicted in Fig. 1a. An optical pulse train (center wavelength $\lambda_0$, width $t_p$, repetition rate $f_{rep}$) is split into two paths, with Path 1 carrying the timing jitter $\Delta t$ to be measured. EOM1, modulated at $f_{rep}/2$, imparts alternating 0 and $\pi$ phase shifts, while EOM2 receives no modulation. When the two pulse trains overlap at the photodetector (PD), they generate a beat note at $f_{rep}/2$, whose envelope encodes $\Delta t$. Envelope detection converts $\Delta t$ to a voltage, yielding $V_1(\Delta t)$ (yellow, Fig. 1b). To enable differential detection, a second identical detection branch is constructed, with the only difference being an additional fixed delay $T_D$ introduced between the two pulse paths before they reach the PD. This yields $V_2(\Delta t)$, which is identical to $V_1(\Delta t)$ but shifted by $T_D$ (red, Fig. 1b). Both curves are peak-normalized, and their difference $V_{out} = V_2 - V_1$ gives the normalized timing characteristic curve (Fig. 1c). This curve consists of multiple sub cycles, each with a zero-crossing. DEST operates at $\Delta t_{TS}$, the zero-crossing with the largest absolute slope. This maximum slope is defined as the normalized timing sensitivity $TS_N$. If the operating point drifts, a fast scan voltage on EOM2 is applied. From the digitized DEST output (via an analog-to-digital converter (ADC)), the four local features (maximum, minimum, average voltage, and polarity) of the sub cycle where the operating point currently resides are extracted. These features uniquely identify the sub cycle and its absolute position on the timing curve, allowing feedback to restore the operating point.

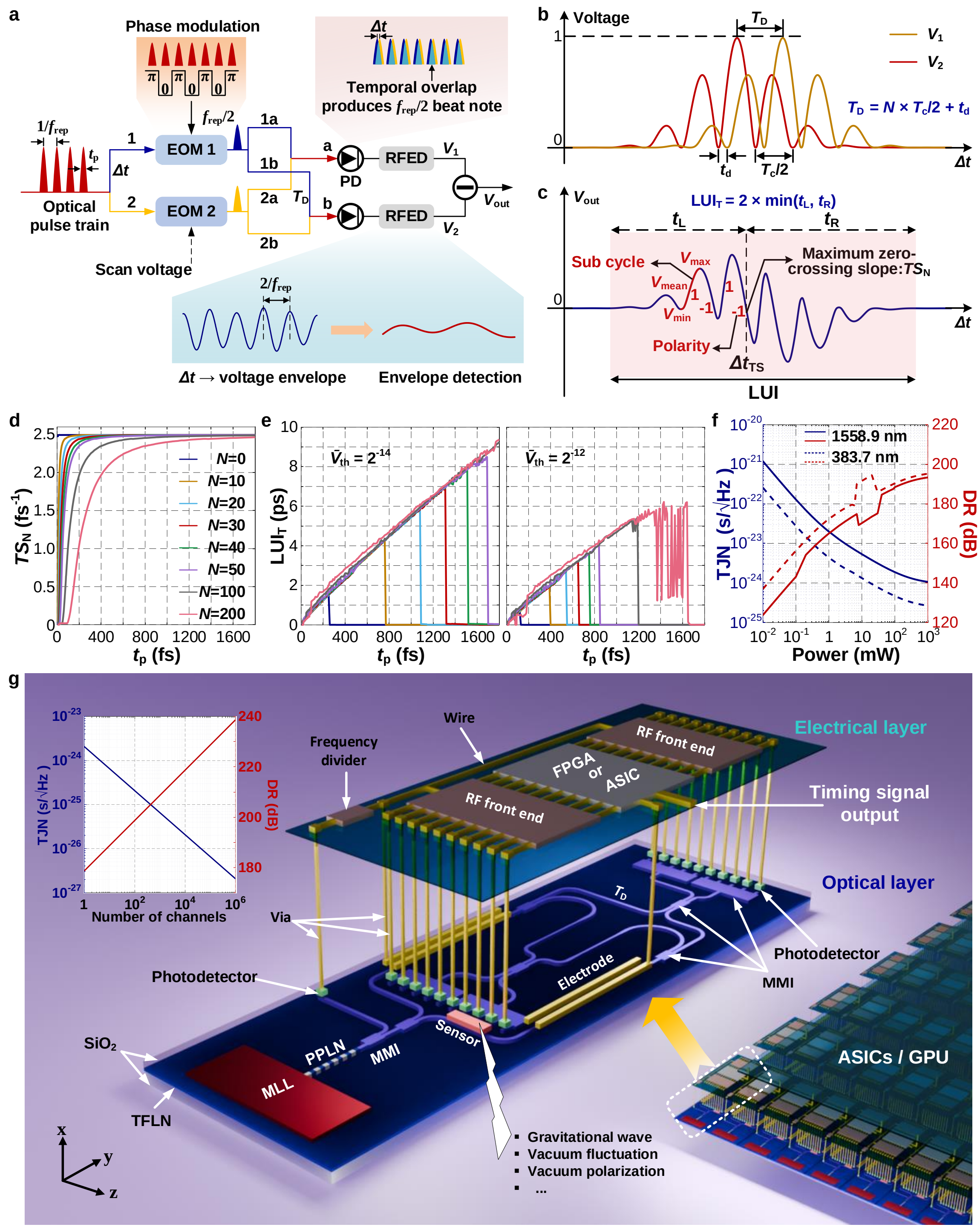

**Fig. 1 Theoretical analysis of DEST. a**, the principle of DEST (EOM, electro-optic modulator; RFED, radio frequency envelope detection). Input pulse: sech($t/\tau$), pulse width $t_p = 1.76\tau$. **b**, $V_1$ and $V_2$ vs. $\Delta t$. **c**, normalized timing characteristic curve $V_{out}$ vs. $\Delta t$. **d**, maximum $TS_N$ (optimized over $t_d \in [0,\ T_c/2]$) vs. $t_p$ for $N$=0, 10, 20, 30, 40, 50, 100, 200 ($T_c$ = 5.2 fs). **e**, $LUI_T$ corresponding to the maximum $TS_N$ in **d**, for normalized resolutions $\bar{V}_{th} = 2^{-14}$ and $2^{-12}$. **f**, Timing jitter noise floor (TJN, left blue) and dynamic range (DR, right red) vs. optical power for $\lambda_0$ = 1558.9 nm ($T_c$=5.2 fs, solid) and 383.7 nm ($T_c$=1.28 fs, dashed). For each data point, $t_d$ is optimized to minimize the TJN. See Supplementary "Noise floor analysis" for details. **g**, conceptual diagram for multi-channel parallel measurement (GPU, graphics processing unit).

In Fig. 1c, the longest unambiguous interval (LUI) is defined as the longest contiguous range of $\Delta t$ within which any two sub cycles can be distinguished from one another using their four local features. To ensure robustness against external perturbations, $\Delta t_{TS}$ should be kept as far as possible from both LUI boundaries. The margin is quantified by $LUI_T$, defined as twice the shorter distance from $\Delta t_{TS}$ to the LUI boundaries (zero if $\Delta t_{TS}$ lies outside). A larger $LUI_T$ means a larger usable measurement range for the DEST (see Supplementary "Timing characteristic curve" for details).

Writing $T_D = N \times T_c / 2 + t_d$ (Fig. 1b), where $T_c$ is the optical carrier period, $N$ is a nonnegative integer and $0 \leq t_d < T_c/2$, a combination of a relatively large $t_p$ and a moderate $N$ can simultaneously provide high $TS_N$ and large $LUI_T$ (see details in Methods). For example, with $t_p = 1.2$ ps, and $N = 200$, we obtain $TS_N = 2.43$ fs$^{-1}$ (Fig. 1d). At a normalized ADC resolution (i.e., the minimum resolvable voltage when the ADC full scale is normalized to unity) $\bar{V}_{th} = 2^{-14}$, the corresponding $LUI_T$ is 6.56 ps (Fig. 1e); even with a coarser resolution $\bar{V}_{th} = 2^{-12}$, $LUI_T$ remains 5.2 ps (Fig. 1e). These values represent over three orders of magnitude improvement in sensitivity and more than an order of magnitude wider unambiguous range than conventional detectors [3-10]. This enables a dynamic range (DR) >160 dB with only 1 mW total optical input power to the two PDs (Fig. 1f). The same figure further reveals that reducing $\lambda_0$ from 1558.9 nm to 383.7 nm lowers the timing jitter noise floor (TJN), highlighting the benefit of operation at shorter wavelengths for enhanced sensitivity.

Fig. 1f also shows that the TJN decreases with increasing total optical power incident on the two PDs. However, the saturation power of low-noise PDs is typically below 2 mW, limiting the minimum achievable timing resolution of a single DEST. Moreover, a free-space or fiber-based implementation would be bulky and susceptible to environmental disturbances, preventing DEST from reaching its full potential. On the other hand, recent advances in thin-film lithium niobate [34] and tantalite [35] waveguide technologies enable monolithic integration of mode-locked lasers (MLL) [36-40], periodically poled lithium niobate (PPLN) [41-45] for frequency quadrupling, electro-optic modulators [34, 35], and photodetectors [46], potentially compressing the entire DEST into a volume of order 1 $cm^3$ (Fig. 2g). This compact form factor facilitates rigorous isolation from the external environment, e.g., through vacuum packaging. Furthermore, compared with discrete implementations, the integrated DEST offers high scalability. As illustrated in Fig. 2g, introducing a 1×$N$ multi-mode Interferometer (MMI) at the output distributes the optical power to multiple on-chip PDs, each operating below saturation while increasing the total input power to the device. Each detector can independently extract timing information, and cross-correlating these signals via a field programmable gate array (FPGA) or an application specific integrated circuit (ASIC) suppresses the timing noise floor. With wafer-scale fabrication, hundreds of DESTs can be produced in a single batch; cross-correlating the outputs of multiple devices can ultimately push the measurement precision to sub-10-rs level while maintaining a dynamic range of 230 dB (upper-left inset of Fig. 2g).

## INTEGRATED DEST DEVICE

As a proof-of-principle demonstration, we fabricated the integrated DEST using the layout shown in Fig. 2a. The chip is built on a 500-nm-thick X-cut lithium-niobate-on-insulator (LNOI) wafer, with an etching depth of 200 nm and a sidewall angle of 75°. The fabrication tolerances for both film thickness and etch depth are within ±10 nm (Fig. 2c). A 700-nm-thick $SiO_2$ cladding is deposited over the lithium niobate (LN) waveguide to improve edge-coupling efficiency and to protect the device from environmental contamination, thereby enhancing its stability.

Optical pulses are coupled into the chip through ports In 1 and In 2, pass through 1.5-cm-long EOM1 and EOM2, respectively, and are then routed through four 1×2 MMIs (Fig. 2d) to implement the splitting and combining scheme of Fig. 1a. A 67.5-µm-long waveguide provides the fixed delay $T_D$. The output signals are collected from Out 1 and Out 2 on both sides of the chip. To further enhance coupling efficiency, double-layer inverse tapers are employed at the edge couplers; Scanning electron microscope (SEM) inspection (Fig. 2e) shows negligible misalignment from lithographic overlay errors.

The electrodes are deposited directly on the $SiO_2$ cladding (Fig. 2f). During fabrication, we evaluated two electrode gaps: 3 µm and 4 µm. The measured half-wave voltage ($V_\pi$) were 3.12 V and 4.68 V (Fig. 2g), respectively. Owing to a ~1-µm overlay misalignment between the LN and electrode layers, the 3-µm-gap devices exhibited 3~4 dB higher insertion loss than the 4-µm-gap ones. We therefore adopted the 4-µm gap for the final design.

The overall chip size is significantly smaller than a coin. A photograph of the packaged device is shown in Fig. 2b, with all four ports fiber-coupled to the outside. The measured fiber-to-fiber insertion losses are approximately 21 dB and 23 dB (Fig. 2h). After excluding the splitting loss of the MMIs, the per-facet coupling loss is estimated to be about 7 dB.

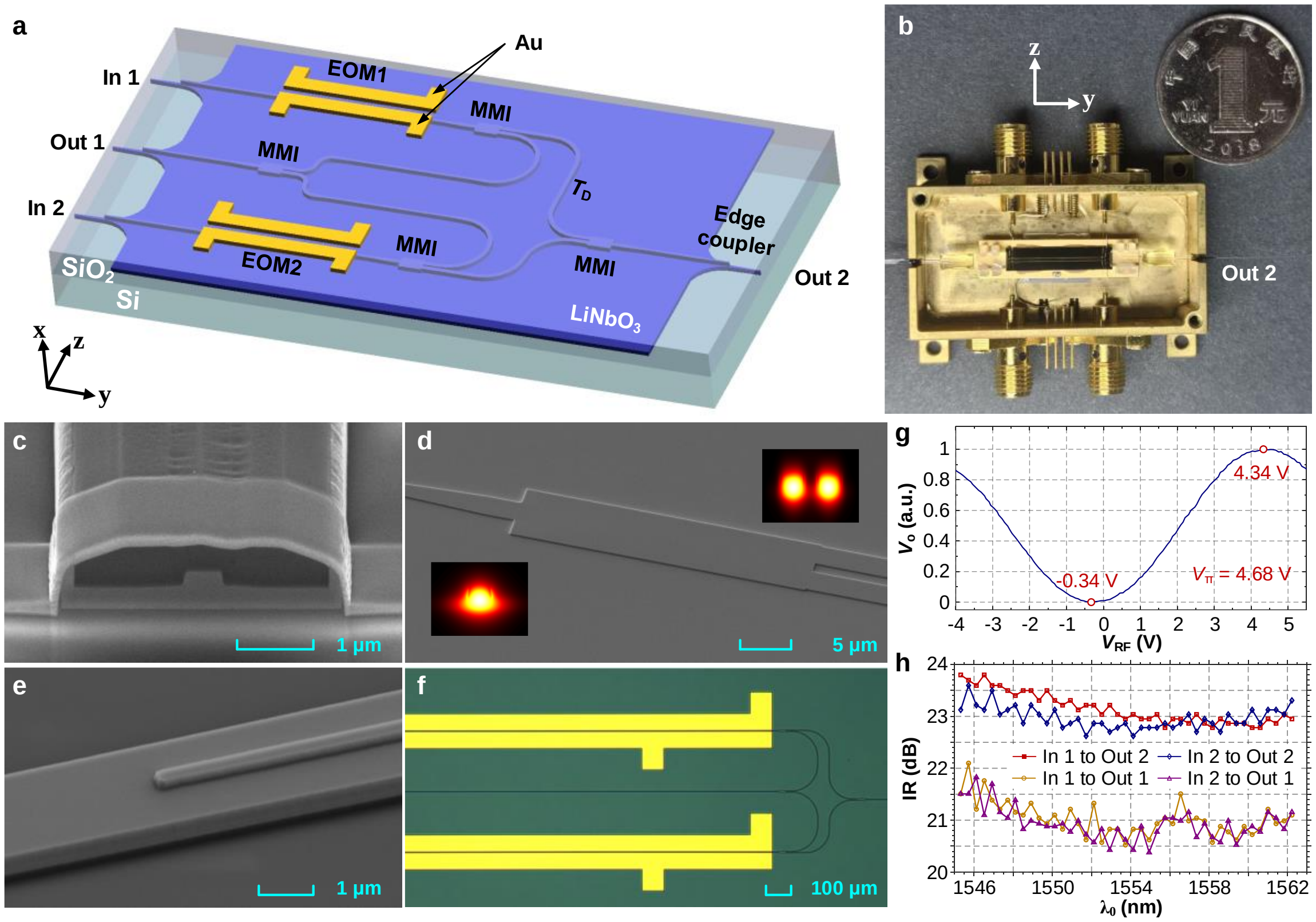

**Fig. 2 Integrated DEST.** **a**, device layout for proof-of-principle demonstration; **b**, photograph of the packaged device; **c**, cross-sectional SEM image; **d**, SEM image of the 1×2 MMI, two insets show the simulated mode field distributions at the input and output ports; **e**, SEM image of the edge coupler; **f**, optical micrograph of the electrodes; **g**, measured $V_\pi$ of the phase modulator with 4-μm electrode gap; **h**, measured fiber-to-fiber insertion loss (IR) of the packaged device.

## TIMING DETECTION NOISE FLOOR MEASUREMENT

We use the setup in Fig. 3a to characterize the timing jitter noise floor of the DEST chip. The Menhir mode-locked laser delivers a pulse train at 1555 nm with 250-fs pulse duration and 216.667-MHz repetition rate. To obtain a sufficiently large LUI, the pulses are first spectrally broadened by self-phase modulation in a single-mode fiber, then filtered by a narrowband optical bandpass filter (OBPF) and amplified, yielding a pulse width of approximately 1200 fs (Fig. 3b). The measured delay of the delay waveguide is $T_D \approx 524$ fs. By tuning the OBPF center wavelength, the ratio $t_d/T_C$ can be adjusted. For example, with the OBPF centered at 1550.8 nm, we have $N = 202$ and $t_d/T_C \approx 0.315$ from $T_D = N \times T_c / 2 + t_d$, corresponding to a normalized timing sensitivity $TS_N \approx 2.4$ fs$^{-1}$, close to the theoretical maximum of 2.43 fs$^{-1}$ for a 1.2-ps pulse (Fig. 1d).

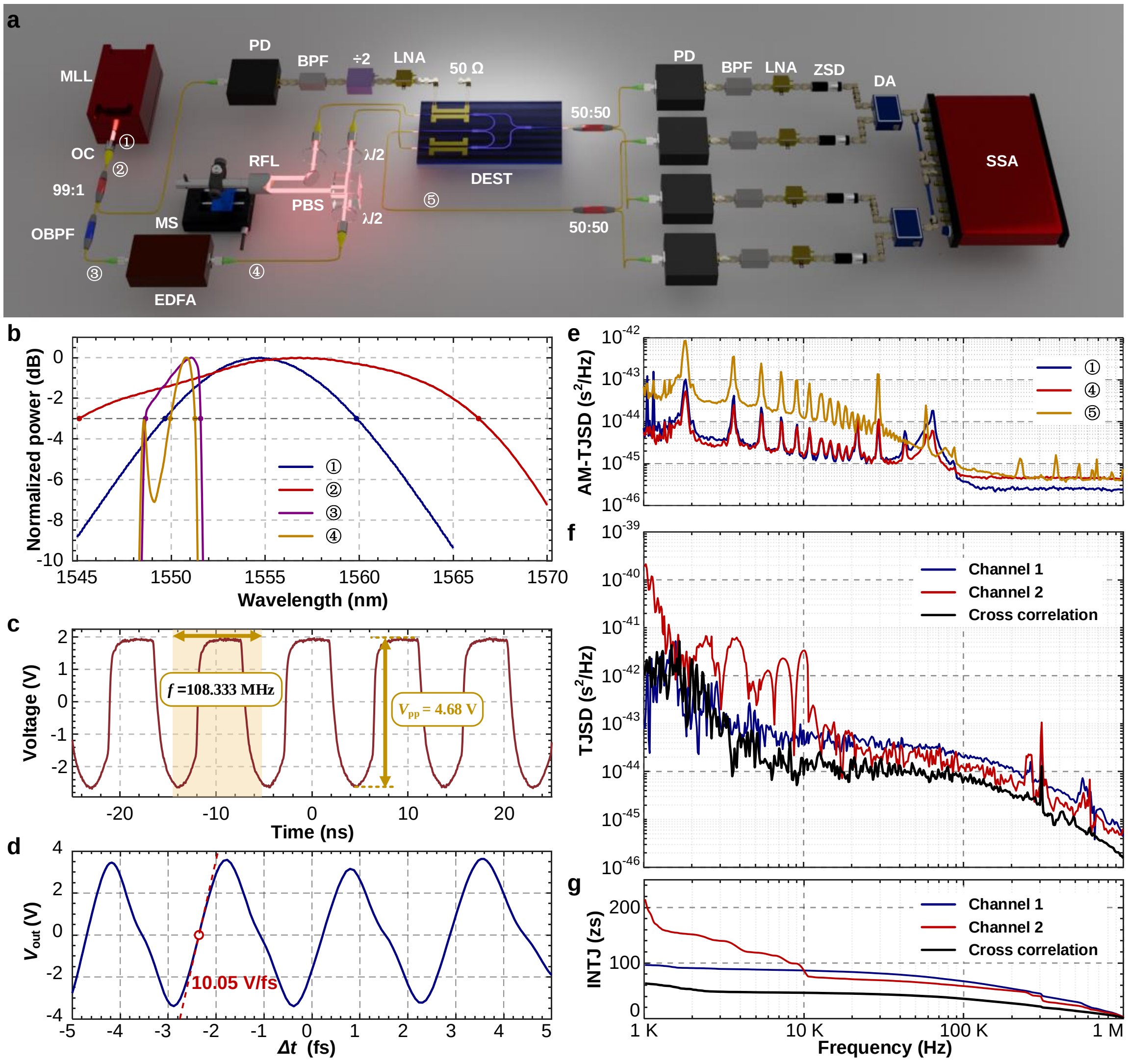

**Fig. 3 Timing noise floor characterization.** **a**, the experimental setup (OC, optical collimator; EDFA, erbium-doped fiber amplifier; MS, motorized stage; RFL, retroreflector; λ/2, half-wave plate; 50 Ω, matched impedance; BPF, band pass filter; ZSD, zero-bias Schottky diode). **b**, Measured optical spectra at positions ①–④ in Fig. 3a; 3-dB bandwidths: 10.184, 21.240, 2.927 and 1.311 nm, respectively. **c**, RF modulation voltage applied to the DEST. **d**, Measured timing characterization curve near $\Delta t_{\mathrm{TS}}$ ($N = 202$, $t_d \approx 0.315\ T_c$). **e**, TJSD induced by amplitude noise, measured at 216.667 MHz (positions ① and ④) and 108.833 MHz (position ⑤). **f**, TJSD of SSA channel 1, channel 2, and their cross-correlation, from three separate acquisitions (SSA measures only one quantity per run). **g**, INTJ from Fig. 3f: 96 zs (channel 1), 215 zs (channel 2), and 63 zs (cross-correlation).

A small fraction of the laser output is tapped by a 99:1 fiber coupler and routed through a PD, a BPF (center frequency 216.667 MHz), a frequency divider by two (÷2), and a low-noise amplifier (LNA) to generate the modulation signal for the DEST (Fig. 3c). The peak-to-peak voltage matches the modulator's $V_\pi$, and the nearly square waveform ensures that the modulation is insensitive to relative timing jitter between the RF signal and the optical pulses.

The amplified optical output is split by a polarization beam splitter (PBS): one path goes directly into the In 2 port of the DEST, while the other passes through a motorized stage (introducing delay $\Delta t$) before entering In 1. The Out 1 and Out 2 outputs are each split by 50:50 couplers into four branches, which are processed by four PDs, BPFs (center frequency 108.333 MHz), LNAs, ZSDs, and two differential amplifiers (DAs) to produce two independent differential outputs. For each differential output, scanning the motorized stage delay yields the timing characteristic curve. With each PD receiving 0.6 mW per pulse train, the measured timing sensitivity is about 10.05 V/fs (Fig. 3d).

Fig. 3e shows the amplitude-noise-equivalent timing jitter spectral density (AM-TJSD) for the bare laser, after amplification, and after the DEST. Below 100 kHz, the DEST noise is dominated by the relative timing jitter between the two input arms; above 100 kHz, the DEST noise floor essentially follows that of the amplified output, indicating that the modulation process introduces no excess noise. Although the amplifier slightly raises the noise floor relative to the bare laser, the lowest noise floor of the amplified output is only $4.33\times10^{-46}$ s²/Hz. After differential operation, this noise floor can be substantially suppressed, ensuring that the TJN of the DEST is not limited by amplitude noise.

The two DA outputs are fed to channel 1 and channel 2 of the signal source analyzer (SSA). The measured timing jitter spectral densities (TJSDs) for each channel are shown in Fig. 3f. After cross-correlation, the noise floor at 1 MHz reaches $1.61\times10^{-46}$ s$^2$/Hz, confirming the viability of the cross-correlation scheme in Fig. 1g. The integrated timing jitter (INTJ) from 1 kHz to 1 MHz after cross-correlation is only 63 zs (Fig. 3g).

## LUI CHARACTERIZATION

The experimental setup in Fig. 4a is used to characterize the LUI of the DEST. To ensure measurement accuracy of the timing characteristic curve, we employ a 24-bit high-precision data acquisition card (DAQ) with a sampling rate of 100 KSa/s. Accordingly, to acquire enough points within each sub cycle of the timing curve, the scan voltage applied to EOM2 must be kept at a frequency on the order of a few hundred hertz. However, at such low modulation frequencies, thin-film lithium niobate modulators suffer from significant relaxation of the electro-optic effect, resulting in greatly reduced modulation efficiency [47]. To circumvent this issue, we apply a sinusoidal voltage at frequency $f_{\mathrm{rep}}+\Delta f$ to EOM2, which equivalently achieves low-frequency modulation of the optical pulses at the difference frequency $\Delta f$. Following the procedure outlined in Fig. 4b (details in Methods), we reconstruct the complete timing characteristic curve shown in Fig. 4c, which comprises 4916 sub cycles.

At a normalized resolution of $\bar{V}_{\mathrm{th}} = 2^{-14}$ (normalized to the peak-to-peak amplitude of the timing characteristic curve), all 4916 sub cycles are mutually distinguishable (Fig. 4d, no entry in the corresponding heatmap equals 4). Since $\Delta t_{\mathrm{TS}}$ is not exactly centered within the acquired range, $\mathrm{LUI_T}$, defined as twice the shorter distance from $\Delta t_{\mathrm{TS}}$ to the two boundaries of the distinguishable interval, is 6.15 ps. This value could be further increased by expanding the acquisition range. As $\bar{V}_{th}$ decreases, the $LUI_T$ gradually shrinks (Fig. 4d and e). At an actual voltage resolution of 0.4 mV, the dynamic range exceeds 155.7 dB (Fig. 4f).

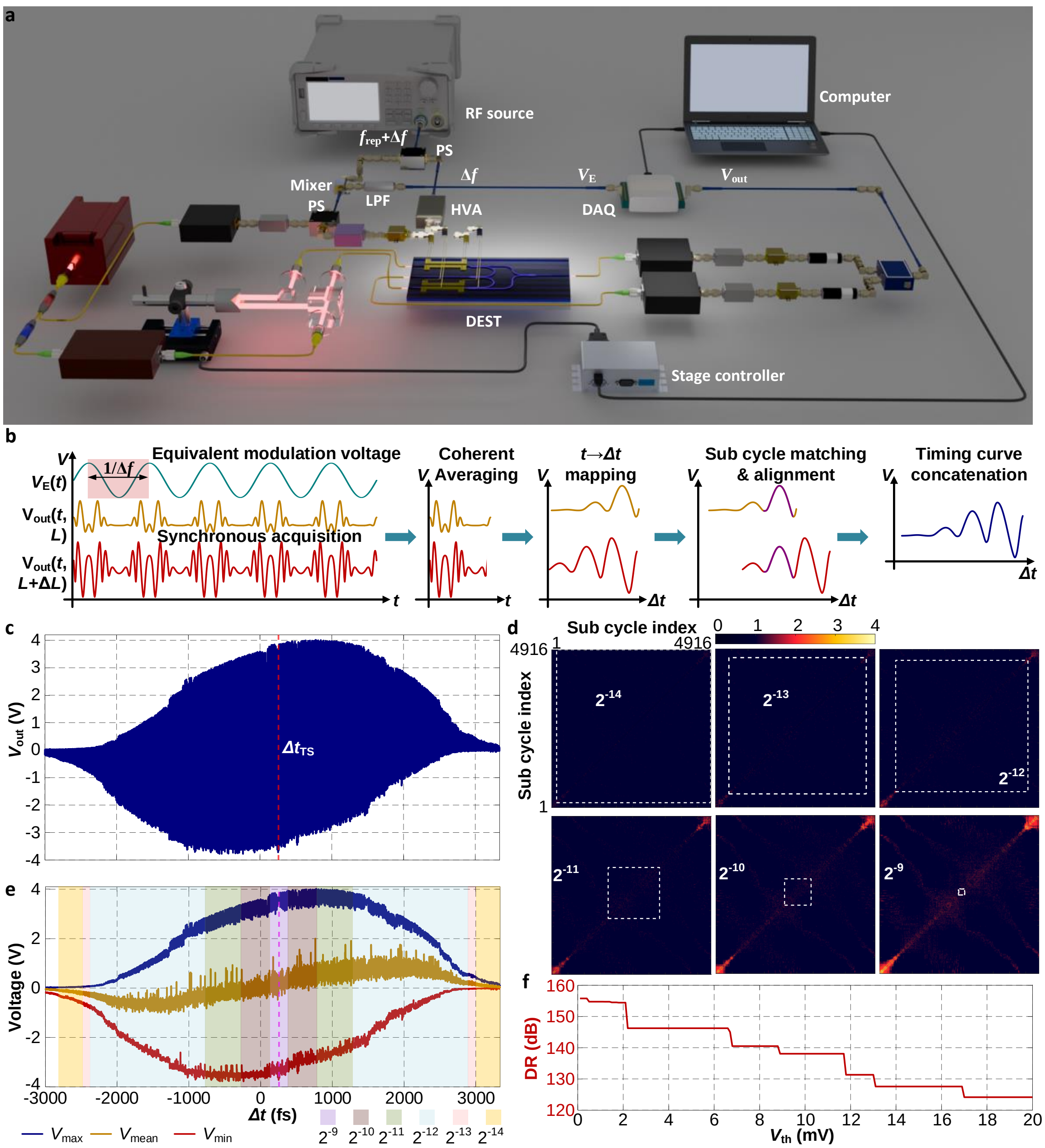

**Fig. 4 LUI characterization.** **a**, the experimental setup (PS, power splitter; LPF, low pass filter; HVA, high voltage amplifier). **b**, Reconstruction of the full timing characterization curve by coherent averaging and segment stitching. **c**, Measured full timing characteristic curve. **d**, Heatmaps of the similarity matrix for different normalized resolutions. Each entry indicates the number of indistinguishable features between two sub cycles; $LUI_T$ is marked by white dashed squares, $LUI_T$ boundaries are constrained by entries equal to 4. For magnified details of the $LUI_T$ boundaries, see Extended Data Fig. 4 (see Supplementary "Timing characteristic curve" for full definition). **e**, Sub-cycle voltage features ($V_{max}$, $V_{mean}$, $V_{min}$) versus Δ$t$. Colored bands mark the $LUI_T$ at different normalized resolutions, with higher resolutions (finer quantization) enclosing lower-resolution intervals. **f**, Dynamic range as a function of threshold voltage, evaluated with the single-channel INTJ @ [1 kHz, 1 MHz] (Fig. 3g) taken as the typical value of 100 zs.

Extended Data Table 1 compares the state-of-the-art timing detection methods. DEST outperforms all previous approaches across multiple metrics: it achieves the lowest real-time timing noise floor (more than three orders of magnitude lower), the smallest integrated timing jitter (nearly two orders of magnitude lower), and the largest $LUI_T$ (more than an order of magnitude wider), all while requiring the lowest optical power (only 1.2 mW per PD). This combination also yields a dynamic range about 66 dB higher than the best prior value.

## DETECTION OF YOCTOSECOND-SCALE PHYSICAL PERTURBATIONS

To further validate the timing resolution of the integrated DEST, we generate and measure extremely weak timing signals using the setup shown in Fig. 5a. The lock-in amplifier provides a single-frequency internal reference signal at $f_{test}$, which is attenuated by 90 dB before being applied to EOM2 of the DEST as the weak timing signal under test. This signal is converted by DEST into a voltage response at the same frequency $f_{test}$ at the DA output. The DA output is then fed into the lock-in amplifier, whose narrow bandwidth extracts the response at $f_{test}$, enabling detection of extremely weak timing signals.

We first verified that the lock-in amplifier can produce voltage signals as weak as 20 nV (see details in Methods). From the measured $V_\pi$ in Fig. 2g, a 20-nV signal corresponds to approximately 11 ys of timing jitter on the electro-optic modulator. To enable long-term monitoring, part of the DA output is split by a PS and fed as an error signal to a proportional-integral (PI) controller, which stabilizes the relative timing jitter between the two pulse trains (at In 1 and In 2 of the DEST) by actuating a piezoelectric transducer (PZT) in the free-space optical path. After lock, the peak-to-peak relative timing jitter is maintained below 20 as (Extended Data Fig. 6), ensuring that the measurement always operates near the maximum slope of the timing characteristic curve.

With the loop locked, the internal reference frequency of the lock-in amplifier is set to 500 kHz, 800 kHz, and 1 MHz. By varying the reference amplitude, different levels of timing jitter are superimposed on EOM2. The resulting DA output amplitudes measured by the lock-in amplifier are shown in Fig. 5b, d, and f. At 1 MHz, the response to 11 ys still remains above the noise floor (Fig. 5f). Furthermore, the modulation voltage applied to EOM2 exhibits a linear relationship with the lock-in-measured DA output amplitude (Fig. 5c, e, g). Since the measurements at larger modulation voltages are reliable, this linearity confirms that the measured 11-ys timing jitter is likewise correct. Moreover, from Extended Data Fig. 9, the noise floor averaged over a 120-s measurement time is estimated to be only 2.89 ys, further supporting the validity of the weak-signal detection.

The current integrated DEST suffers from relatively high insertion loss (Fig. 2h),

primarily arising from ~7 dB per-facet edge-coupling loss and the ~3 dB loss of the output 2×1 MMIs. In future designs, replacing the 2×1 MMIs with 2×2 MMIs at the output stage, combined with improved packaging techniques, could reduce the total insertion loss by approximately 13 dB. This improvement would lower the noise floor shown in Extended Data Fig. 9 to below 650 rs.

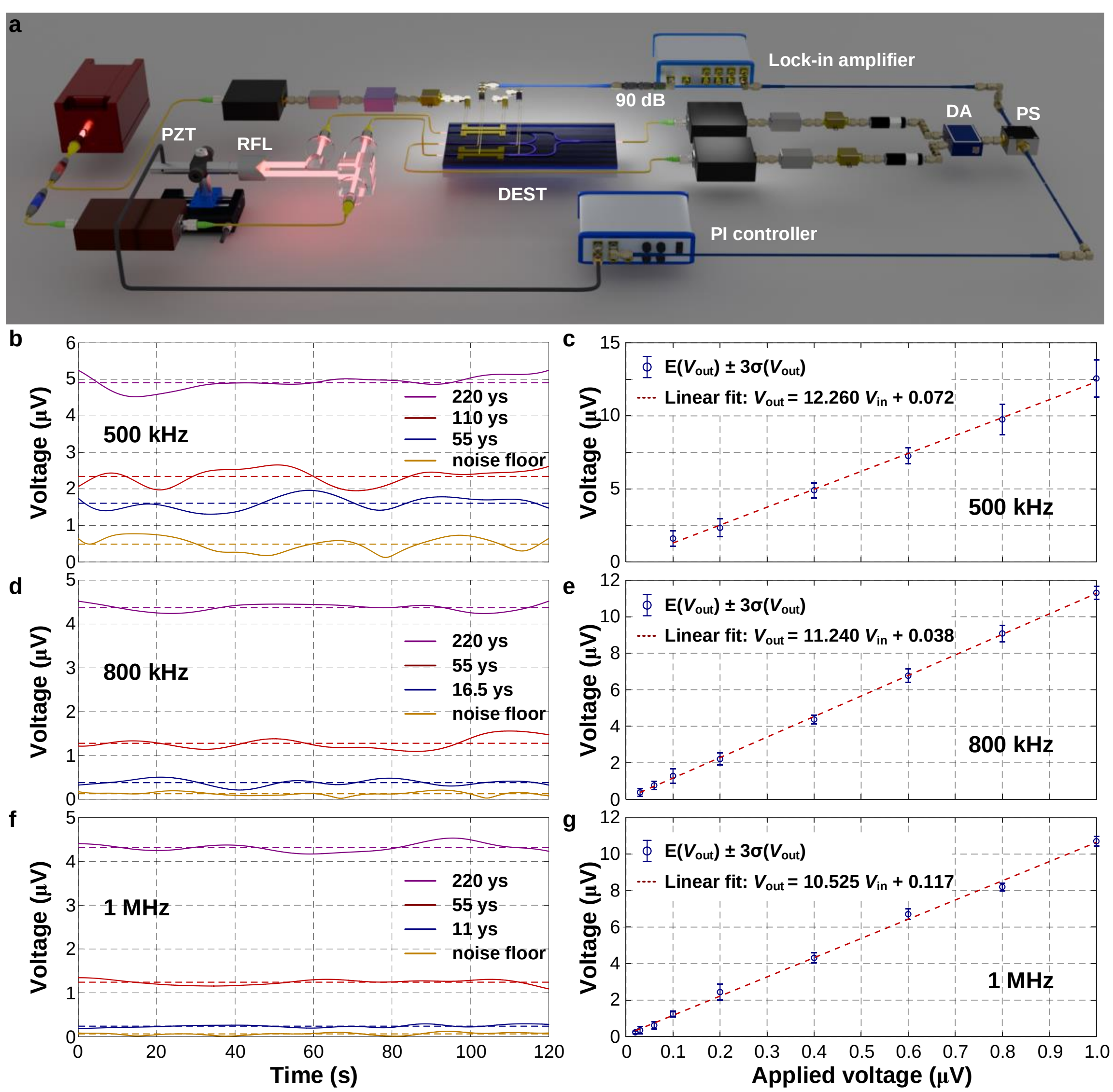

**Fig. 5 Measurement of extremely weak timing signals. a**, the experimental setup. **b, d and f**, Voltage amplitude traces measured by the lock-in amplifier for different applied timing jitter amplitudes at 500 kHz (b), 800 kHz (d) and 1 MHz (f), together with the noise floor (no modulation input). **c, e, and g**, Lock-in amplifier measurements (mean ± 3σ) versus EOM2 modulation amplitude at 500 kHz (c), 800 kHz (e) and 1 MHz (g), with a linear fit of the mean output voltage against input amplitude for each case.

## DISCUSSION

Beyond its ys-precision resolution and >155 dB dynamic range, integrated DEST offers several additional advantages. One important advantage is that the requirements for on-chip pulse sources are greatly relaxed. This is achieved in two ways. First, the incident pulses do not need to maintain their transform-limited pulse width: based on Eq. (S8) in the Supplementary Information and Parseval's theorem, any common chirp between the two pulses is automatically canceled in the photodetection process, leaving the timing measurement unaffected. Second, as shown in Fig. 1d and e, achieving simultaneously high $TS_N$ and large $LUI_T$ calls for picosecond pulses—rather than the hundred-femtosecond or sub-ten-femtosecond pulses required by conventional timing detectors. Owing to these two features, the timing source can be generated by either a TFLN-based pulse generator [48] or a Kerr frequency comb [49, 50] with the background pump light filtered out.

The second advantage of DEST is its strong immunity to various noise sources that commonly plague conventional timing detectors. The TJSD in Fig. 3f is measured above 1 kHz; below that, the noise is dominated by thermal fluctuations in the fiber connections of the packaged device. This low-frequency noise can be substantially suppressed by monolithic integration (Fig. 2g), which shortens the optical path and allows precise temperature stabilization through proper packaging.

Compared with direct laser interferometry, DEST upconverts the timing jitter to the $f_{rep}/2$ band, physically isolating the additive component of laser amplitude noise that would otherwise corrupt the signal in baseband detection. In DEST, the same amplitude noise becomes multiplicative and is effectively suppressed by the differential operation.

The present implementation uses ZSD detection as a low-cost proof-of-principle approach. In the future, direct digitization of the $f_{rep}/2$ RF timing signal, followed by digital envelope extraction, could eliminate all baseband analog circuitry. This would circumvent flicker noise and extend ys-precision to low frequencies, enabling the detection of extremely weak signals in this band—such as those expected from gravitational waves.

The third advantage is that massive array integration does not compromise the real-time nature of DEST. Let $X_i$ denote the Fourier transform spectrum of the $i$-th channel signal. The averaged pairwise cross-correlation over all channel pairs can be expressed in the frequency domain, up to an irrelevant factor, as

$$P_{\text{c-c}} = \left|\sum_{i=1}^{M} X_i\right|^2 - \sum_{i=1}^{M} \left|X_i\right|^2 \quad (1)$$

where $M$ is the total number of channels. This formulation enables highly efficient computation using parallel architectures. As shown in the Supplementary Information (Computational resource estimation), the output latency is limited only by the sampling period and the fast Fourier transform (FFT) length, making real-time operation feasible even for massive arrays. Based on recently demonstrated integrated MLLs [36-40] and PPLN waveguides [41-45], the footprint of the fully on-chip DEST in Fig. 2g is expected to be within $20 \times 4$ mm$^2$. Assuming a single chip supports 16 parallel detection channels, a 1m$^2$ area could accommodate up to $2 \times 10^5$ channels. For such an array, with a sampling rate of 1 GSa/s and an FFT size of 1024, the timing output latency is only 3.156 μs (see Supplementary "Computational resource estimation" for detailed calculations). Furthermore, when the operating point deviates from $\Delta t_{TS}$, a scan voltage applied to EOM2 for only about 50 ns (at 1 GSa/s with $t_p$ = 1.2 ps) is sufficient to extract the sub cycle features needed to determine the current sub cycle position (Supplementary: Sampling points requirements). This rapid response ensures that the DEST array can not only measure but also actively control timing jitter in real time, even at the scale of hundreds of thousands of channels.

In Fig. 4a, the motorized stage (millimeter range) ensures a large $LUI_T$; in the fully integrated version (Fig. 2g), it can be replaced by an electrically controlled tunable delay line on TFLN [51]. The sensor in Fig. 2g can be implemented by integrated structures such as surface-plasmon-resonance sensors [52], slot-waveguide [53], or cantilevers [54]. Alternatively, gratings combined with micro lenses can collimate the optical field into free space to probe weak effects such as gravitational waves, vacuum polarization, and vacuum fluctuations. Through parallel cross-correlation on a 1-$m^2$ integrated DEST array, the TJN could be reduced below 4.6 rs/√Hz, corresponding to a spatial strain sensitivity of $1.39\times10^{-18}$ m/$Hz^{-1/2}$. This surpasses the minimum sensitivity of Advanced LIGO ($3.36\times10^{-18}$ m/$Hz^{-1/2}$ for 1200 km optical path length) [55], currently the most sensitive detector developed by humanity.

## METHODS

**The choice of $T_D$ and $t_p$.** In designing the DEST, careful selection of the time delay $T_D = N \times T_c / 2 + t_d$ and the pulse width $t_p$ is essential for optimizing $TS_N$, LUI and $LUI_T$. As $T_D$ varied, both $TS_N$ and LUI exhibit quasi-periodic behavior with period $T_c/2$, but their maxima occur at different values of $t_d$ within each period (Extended Data Fig. 1b). In particular, the maximum $TS_N$ does not coincide with the peak-to-peak maximum of the timing curve (Extended Data Fig. 1a), whereas the LUI generally peaks at $t_d = T_c/4$ (Extended Data Fig. 1c). As $N$ increases, $TS_N$ decreases monotonically, while the LUI first grows and then drops sharply beyond a certain threshold (Extended Data Fig. 1d). For a fixed pulse width $t_p$, a smaller $N$ yields a larger $TS_N$; larger $t_p$ slows the decline of $TS_N$ with increasing $N$ (Fig. 1d). For a fixed $N$, $LUI_T$ increases with $t_p$, but once $t_p$ exceeds a threshold, the LUI suddenly shrinks—pushing $\Delta t_{TS}$ toward the LUI edge—and $LUI_T$ drops sharply (Extended Data Fig. 2). This threshold rises with $N$ (Fig. 1e). Therefore, a relatively large $t_p$ combined with a moderate $N$ can simultaneously provide high $TS_N$ and large $LUI_T$.

**The fabrication of integrated DEST.** A 200-nm-thick chromium hard mask is first deposited on the LNOI wafer, followed by spin-coating of hydrogen silsesquioxane (HSQ) resist. After electron-beam lithography (EBL), inductively coupled plasma (ICP) etching sequentially removes the hard mask and 200 nm of the LN film, defining the ridge waveguide structures. A second EBL step with an 800-nm-thick Si hard mask and ICP etching removes 300 nm of the LN film at the chip edges to form the edge couplers. Next, a 0.7-µm-thick $SiO_2$ cladding is deposited by plasma enhanced chemical vapor deposition (PECVD). Electrodes are then defined by ultraviolet lithography, followed by gold deposition and lift-off.

The edge couplers terminate in 200-µm-long polishing structures with rounded ends (Extended Data Fig. 3), designed to prevent fracture during processing. After chip fabrication, these 200-µm end sections are polished away on both facets, yielding the final edge couplers shown in Fig. 2e.

**Reconstruction of the full timing characterization curve**. As shown in Fig. 4a, to reconstruct the complete timing curve, an RF source generates a sinusoidal signal at frequency $f_{rep} + \Delta f$, which is then split by a PS into two paths. One path is amplified by an HVA and directly applied to EOM2 of the DEST. The peak-to-peak amplitude of this modulation signal is set to at least three times the half-wave voltage $V_\pi$, ensuring that the DEST output $V_{out}$ sweeps across at least six sub cycles during each modulation period. The other path is mixed with the laser repetition rate $f_{rep}$ and low-pass filtered to produce a low-frequency signal $V_E$ at the difference frequency $\Delta f$. This signal $V_E$ represents the equivalent modulation voltage actually experienced by the optical pulses in EOM2.

We then synchronously acquire both $V_E$ and the DEST output $V_{out}$ using a DAQ. Since $V_E$ is a periodic sinusoidal signal, each of its samples corresponds to a specific phase; consequently, the synchronously acquired $V_{out}$ also becomes periodic (Fig. 4b). However, temperature drift during acquisition can cause $V_{out}$ at the same phase of $V_E$ to vary from one modulation period to another. To correct for this, we apply a time shift to align the extrema of $V_{out}$ across different periods—ensuring that the maximum (or minimum) of $V_{out}$ consistently corresponds to the same phase of $V_E$. After this alignment, we perform coherent averaging by averaging all $V_{out}$ samples corresponding to the same phase of $V_E$ over multiple periods. This yields a precise relationship between the equivalent modulation voltage $V_E$ and the DEST output $V_{out}$ within a single period. Finally, using the measured half-wave voltage $V_\pi$, we convert each value of $V_E$ into the corresponding timing jitter $\Delta t$ (i.e., $t \rightarrow \Delta t$ mapping), thereby transforming the voltage–voltage relationship into a segment of the $\Delta t$-$V_{out}$ curve.

Next, we move the motorized stage by a small step $\Delta L$ and repeat the above procedure to obtain an adjacent segment of the $\Delta t$-$V_{out}$ curve. The step size $\Delta L$ is chosen such that the two segments share at least four overlapping sub cycles, while also containing at least two distinct non-overlapping sub cycles. Using the overlapping region as a reference, we stitch the two segments together. By repeating this process—moving the stage incrementally and stitching successive overlapping segments—we reconstruct the entire timing curve over the full range of interest.

**PI controller locking**. To ensure robust locking in Extended Data Fig. 6, given the narrow linear range of the timing-detector error signal, several measures are implemented. The detector output is first attenuated by 15 dB before entering the PI controller. The PZT actuates a retroreflector (RFL) mounted on a V-groove (see inset of Extended Data Fig. 6). The V-groove supports the RFL to counteract the shear force exerted on the PZT by the weight of the RFL and its mounting screw, while also providing lateral damping to prevent overshoot beyond the error-signal linear region. The PI controller has a lock bandwidth of approximately 30 Hz, which is sufficient to suppress low-frequency thermal drift while avoiding the injection of high-frequency noise that could interfere with weak-timing-signal measurements.

**Lock-in amplifier**. A Zurich Instruments MFLI-DC-5M lock-in amplifier is used in the experiment, with an input voltage noise of 2.5 nV/√Hz above 1 kHz. The internal reference signal is phase-shifted and amplified, then output from the Signal Output port and attenuated by 90 dB to generate the weak modulation signal for the DEST. The DEST output (under PI lock) is received through the Signal Input port. The lock-in performs synchronous multiplication demodulation between the internal reference and the input signal, followed by low-pass filtering to extract the DC output component, which contains the timing signal of interest. An 8-th order digital low-pass filter with a bandwidth of 10 mHz is used.

To minimize crosstalk between the Signal Output and Signal Input ports, the RF cables from the two ports are kept more than 20 cm apart and are arranged to avoid parallel routing. After each adjustment of the output amplitude or frequency via the control software, a 2-minute settling time is allowed before data acquisition to ensure the filter output reaches steady state.

To verify the lock-in amplifier's ability to generate weak signals. The internal reference provides 1-MHz signals with amplitudes of 20 μV and 30 μV, which are attenuated by three cascaded 30-dB attenuators, and directly connected to the lock-in input. The internal reference phase is set to 0 and the X component of the Signal Input is recorded. Compared with the no-input case, the mean readings increase by approximately 20.303 nV and 30.926 nV, respectively (Extended Data Fig. 5), confirming that the lock-in can indeed produce very weak voltage signals.

For the timing-signal measurements in Fig. 5b–g, the R component of the Signal Input is recorded. All data are saved automatically by the control software.

## Data availability

The data that support the findings of this study are available from the corresponding authors on request.

**Acknowledgements** This work was supported by the National Natural Science Foundation of China (Grant No. 62575209), National Key R&D Program of China (Grant No. 2021YFC2201902), Chinese Academy of Sciences Project for Young Scientists in Basic Research (YSBR-059) and National Natural Science Foundation of China (U25A20515).

**Author contributions** M.X. conceived the concept. J.Y. designed and tested the chip. J.Y. and T.W. performed the experiment and analyzed the data. B.W. assisted with the chip test. Y.S., D.P., Z.C. and S.F assisted with the experiments. M.C. packaged the chip. M.C., G.W., M.Z. and M.X. initiated and supervised the project. All authors prepared the manuscript.

**Competing interests** The authors declare that they have no conflict of interest.

**Additional information**

**Correspondence and requests for materials** should be addressed to M.X.

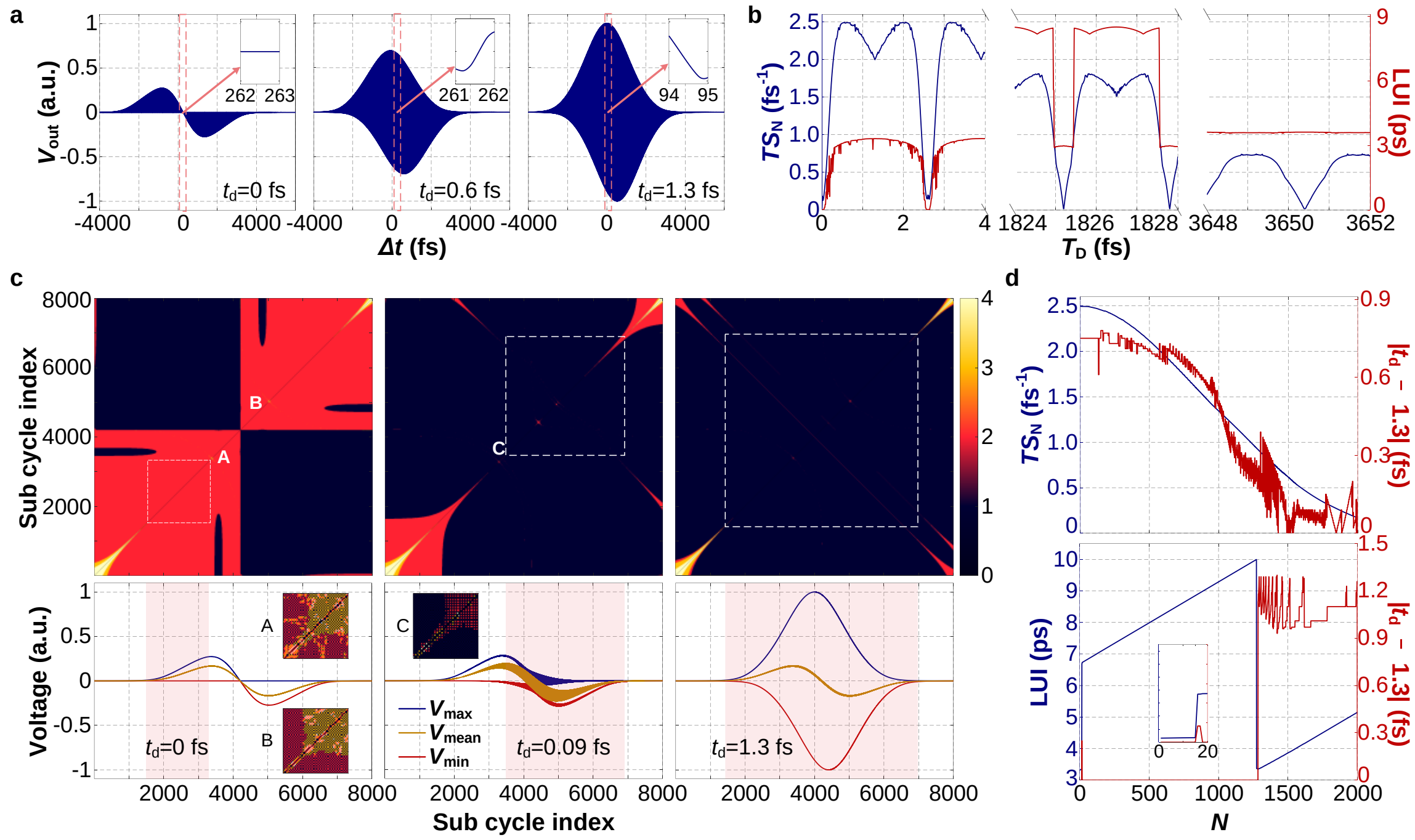

**Extended Data Fig. 1 Simulation results of $TS_N$ and LUI**. $T_c$= 5.2 fs for all calculations. **a**, Normalized timing characterization curves for $N$=202 at $t_d$ = 0, 0.6, 1.3 fs, $t_p$=1.2 ps. $TS_N$: 0.0063, 2.4191 and 1.9494 fs$^{-1}$. Insets: zoom near $TS_N$. **b**, $TS_N$ (left blue) and LUI (right red) vs. $T_D$. **c**, Similarity matrix heatmaps and sub-cycle voltage features ($V_{max}$, $V_{mean}$, $V_{min}$) vs. sub-cycle index for $N$=202, $t_p$=1.2 ps, $\bar{V}_{th} = 2^{-14}$ at $t_d$ = 0, 0.09, 1.3 fs. LUI: white dashed squares (heatmaps) and light red rectangles (curves). At $t_d$=0 and 0.09 fs, entries=4 in regions A-C limit LUI. At $t_d$=1.3 fs ($T_c$/4), all three voltage features show largest excursions, yielding the longest LUI. **d**, Local maxima of $TS_N$ (top) and LUI (bottom) vs. $N$ (left axes), with |$t_d$-1.3 fs| at each maximum (right axes). For each $N$, the maximum is found by scanning $t_d \in [0, T_c/2]$.

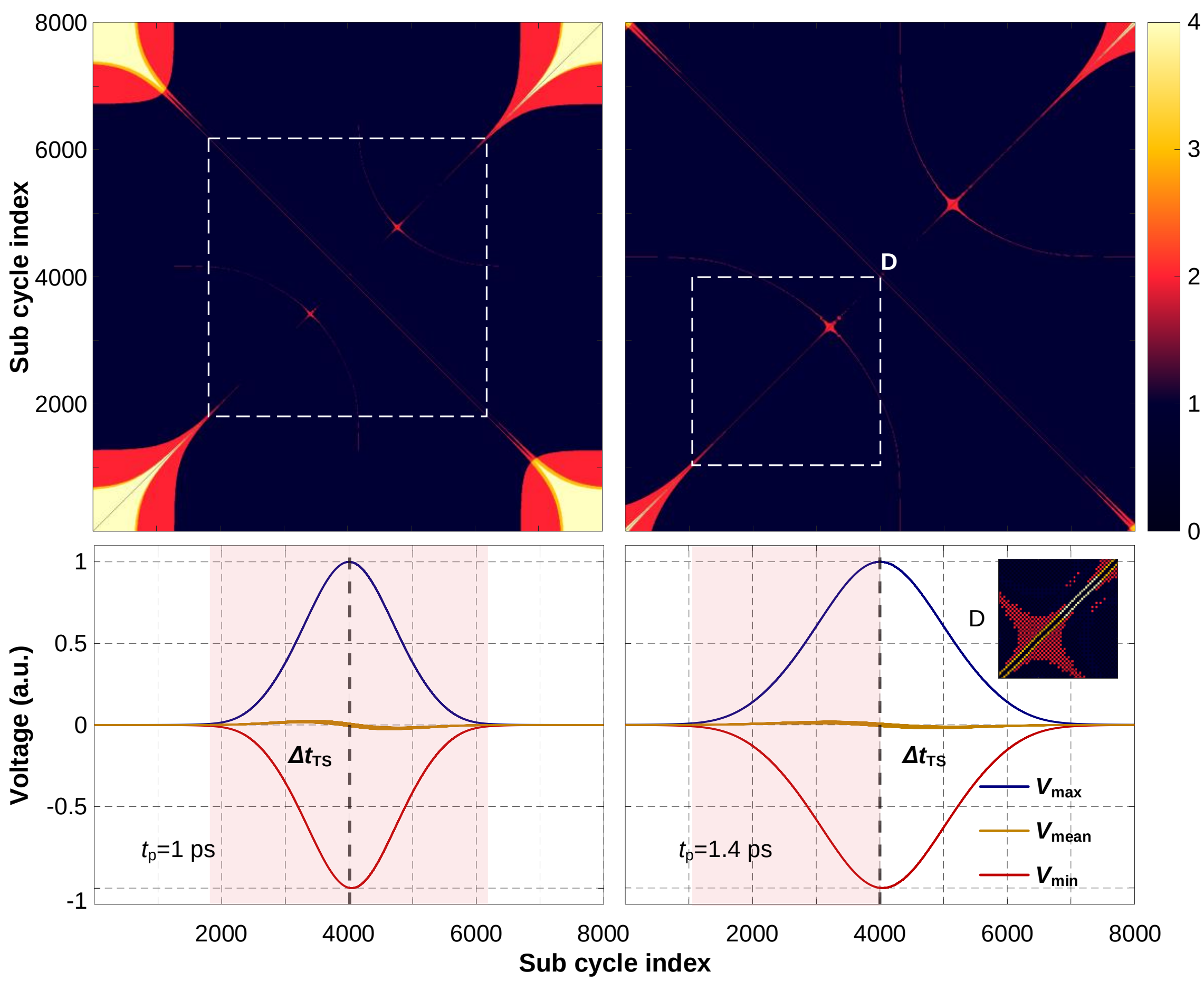

**Extended Data Fig. 2 Effect of pulse width $t_p$ on $LUI_T$**. For $T_c$ = 5.2 fs, $N$ = 20, $t_p$ = 1 ps (left) and $t_p$ = 1.4 ps (right). Top: similarity matrix heatmaps with LUI (white dashed squares). Bottom: sub-cycle voltage features with LUI (light red rectangles) and $\Delta t_{TS}$ (black dashed). $t_p$ =1 ps, $\Delta t_{TS}$ near center → large $LUI_T$; $t_p$=1.4 ps, $\Delta t_{TS}$ near right edge → small $LUI_T$.

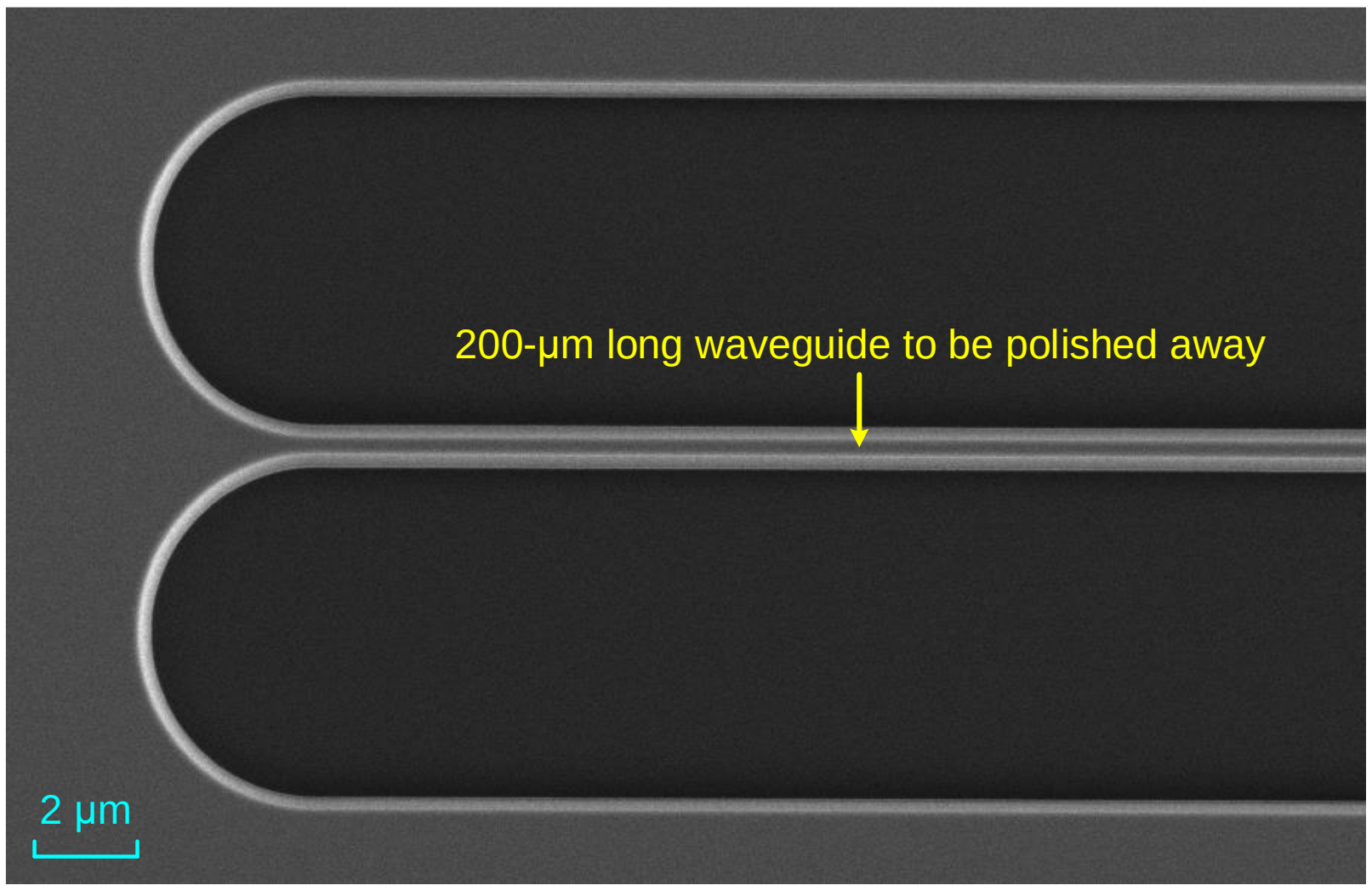

**Extended Data Fig. 3** SEM image of the polishing structure at the chip edge.

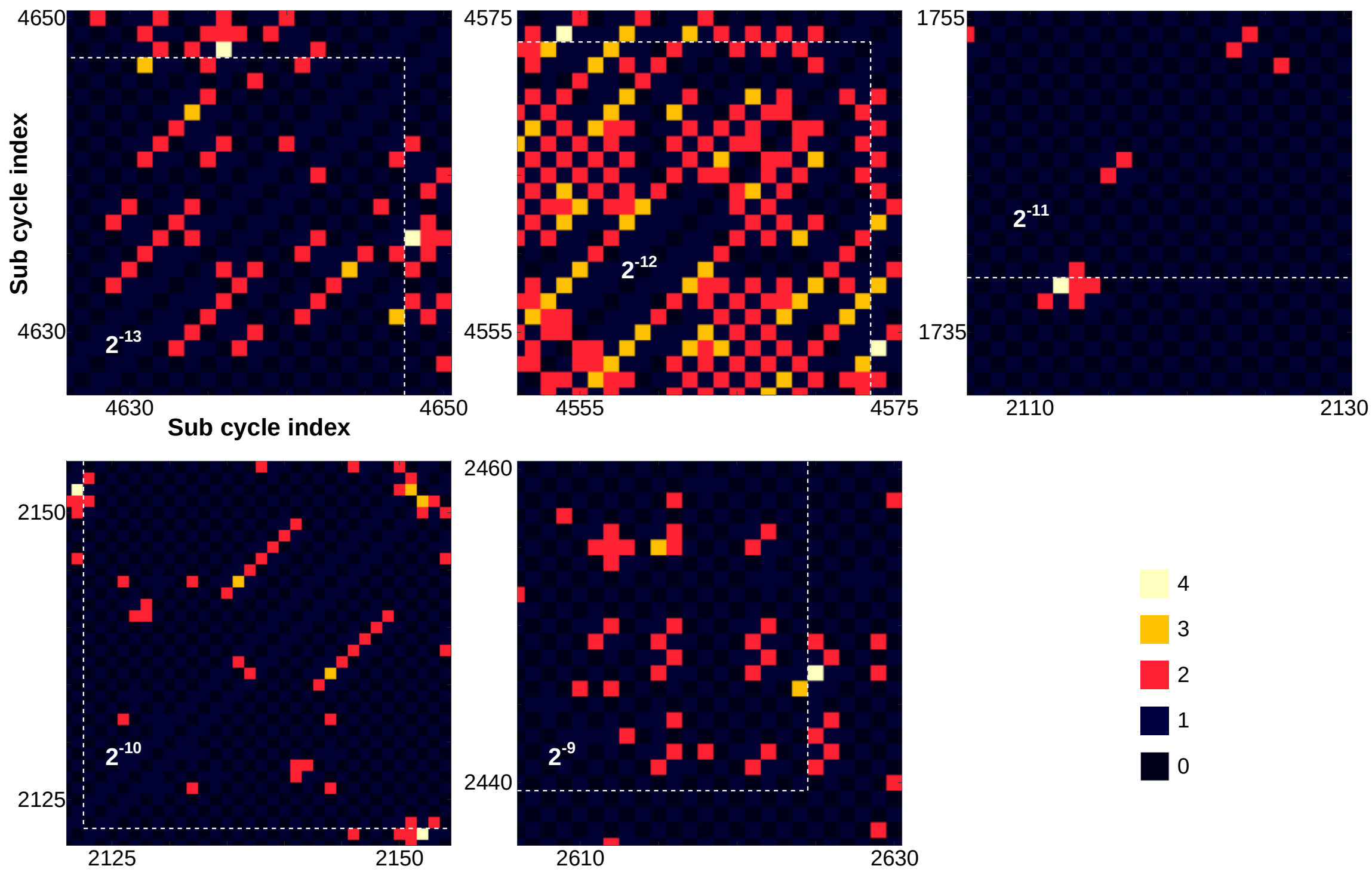

**Extended Data Fig. 4** Zoomed-in view of the heatmaps in Fig. 4d, revealing the entries equal to 4 that define the $LUI_T$ boundaries.

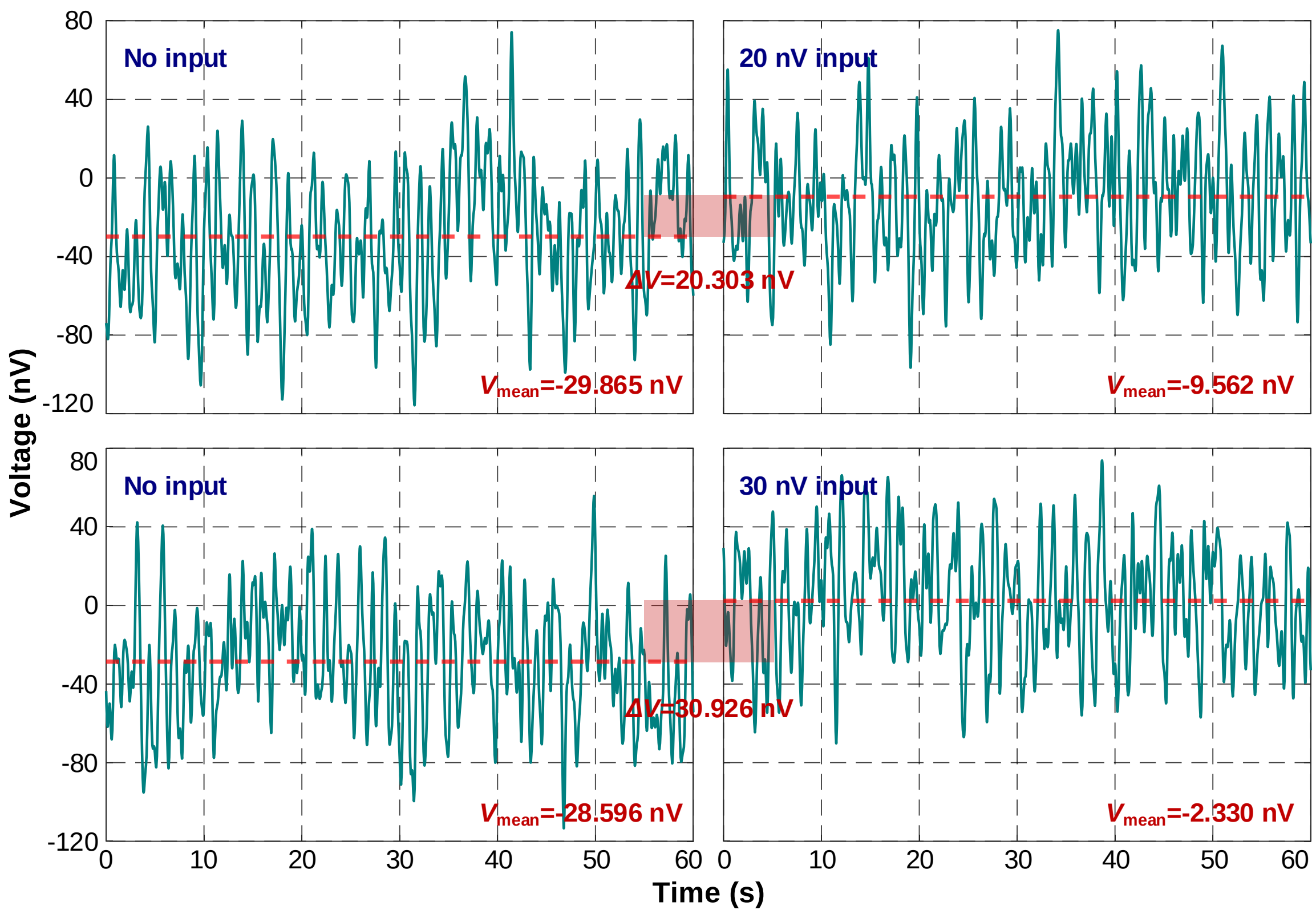

**Extended Data Fig. 5** Measured lock-in amplifier response to 20 nV (top) and 30 nV (bottom) inputs.

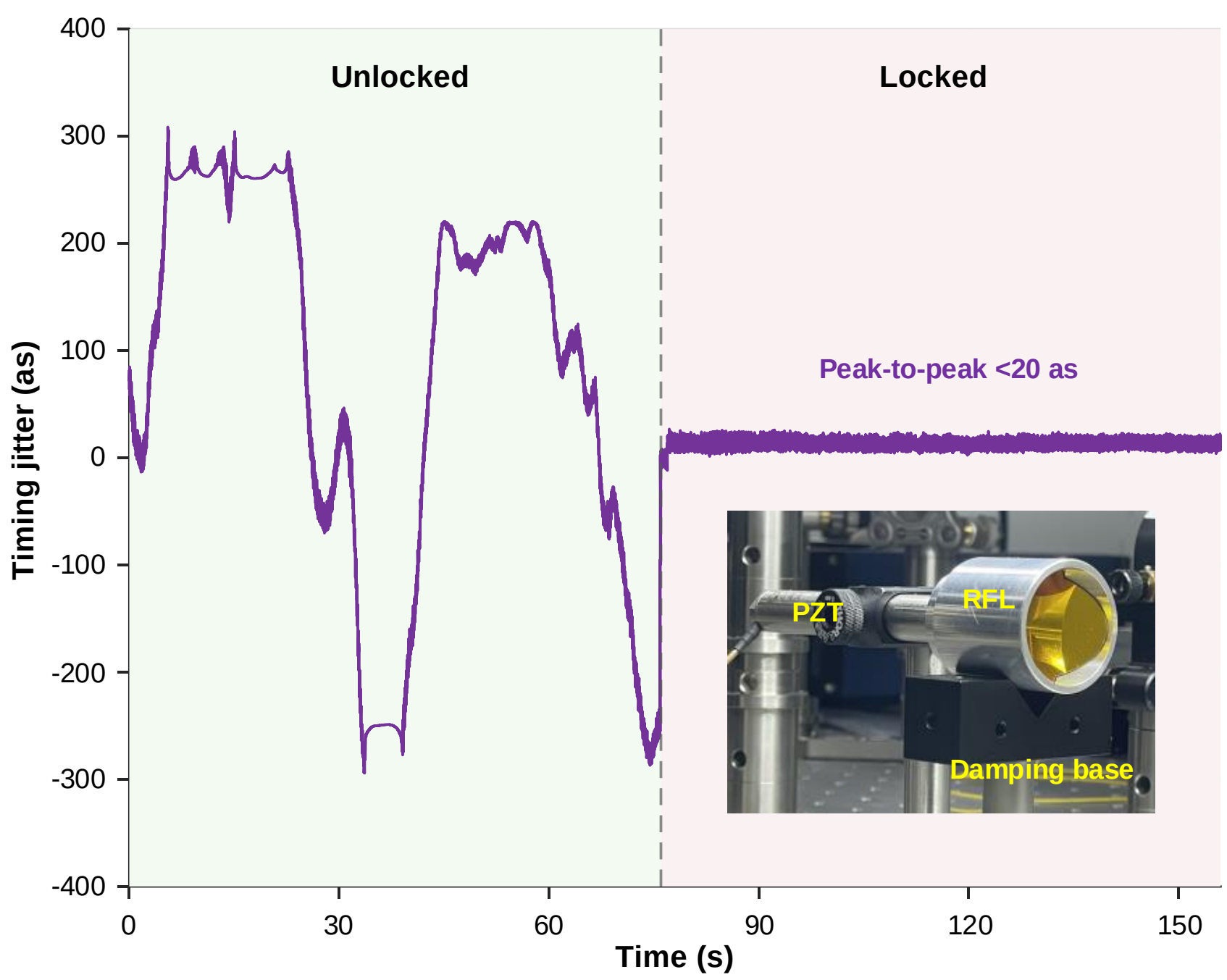

**Extended Data Fig. 6** Relative timing jitter before and after PI lock (actuator photo in inset).

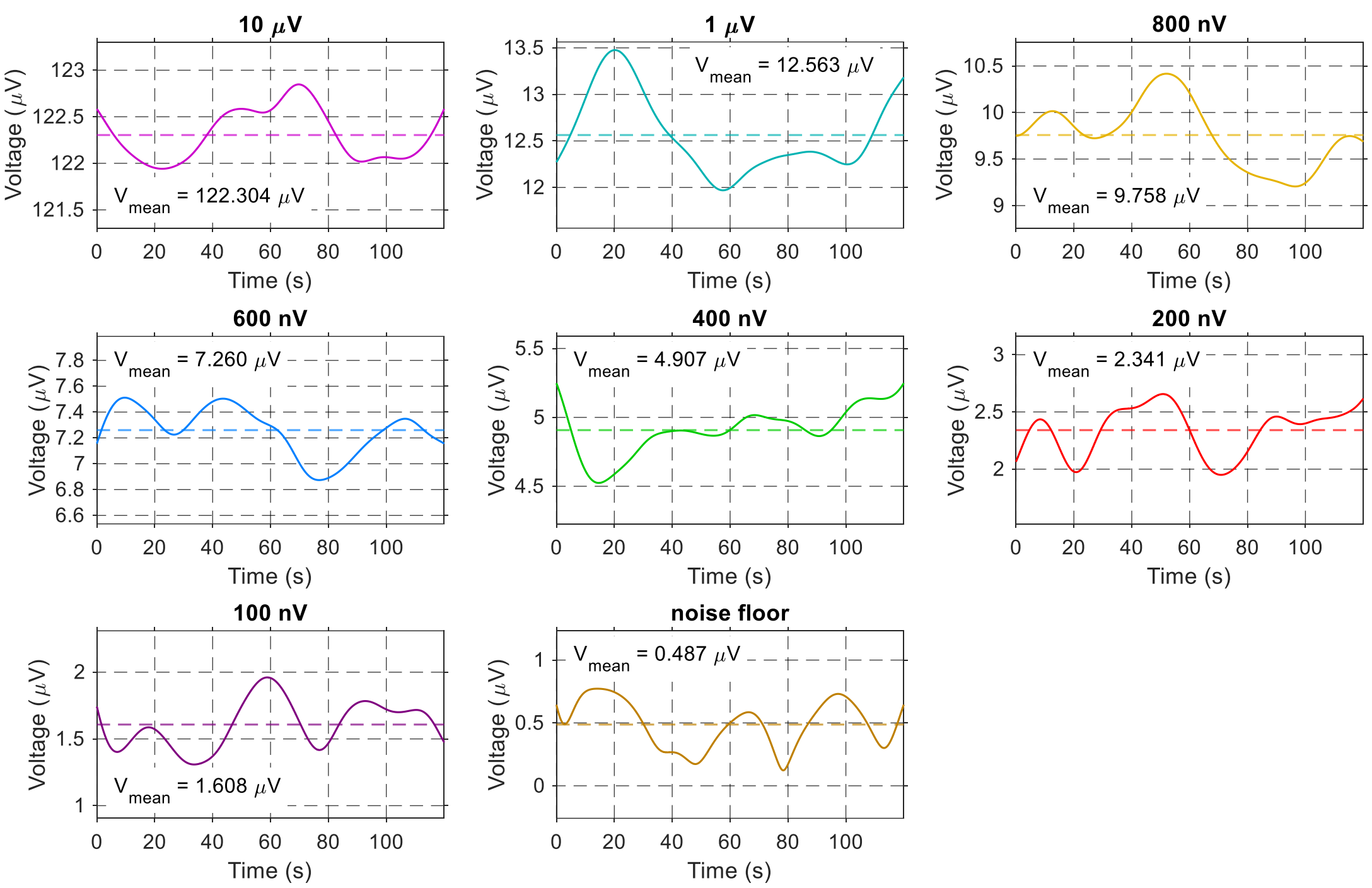

**Extended Data Fig. 7** Full raw measurement data for the 500-kHz extremely weak timing signal; solid lines represent experimental traces and dashed lines denote the averaged values.

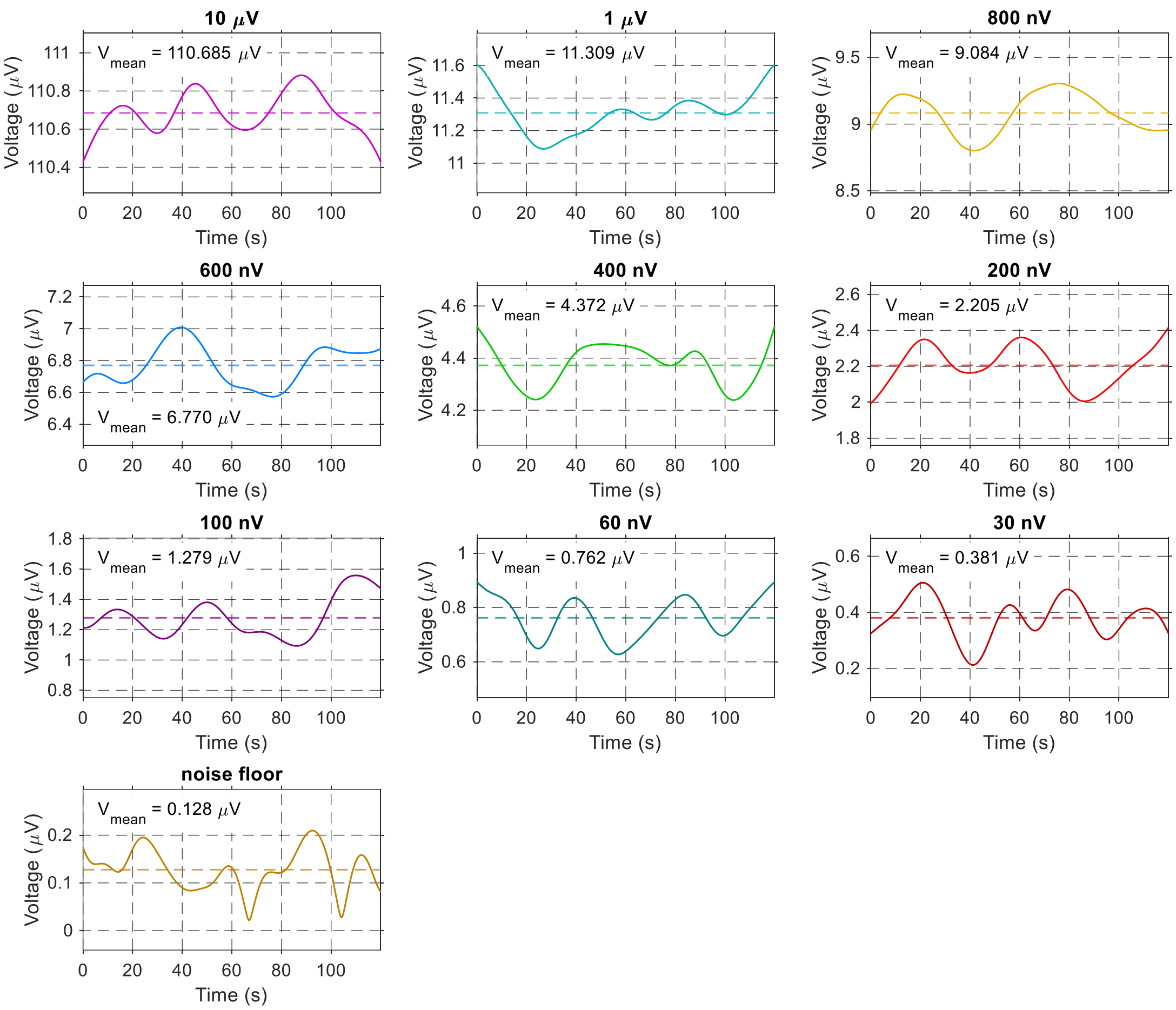

**Extended Data Fig. 8** Full raw measurement data for the 800-kHz extremely weak timing signal; solid lines represent experimental traces and dashed lines denote the averaged values.

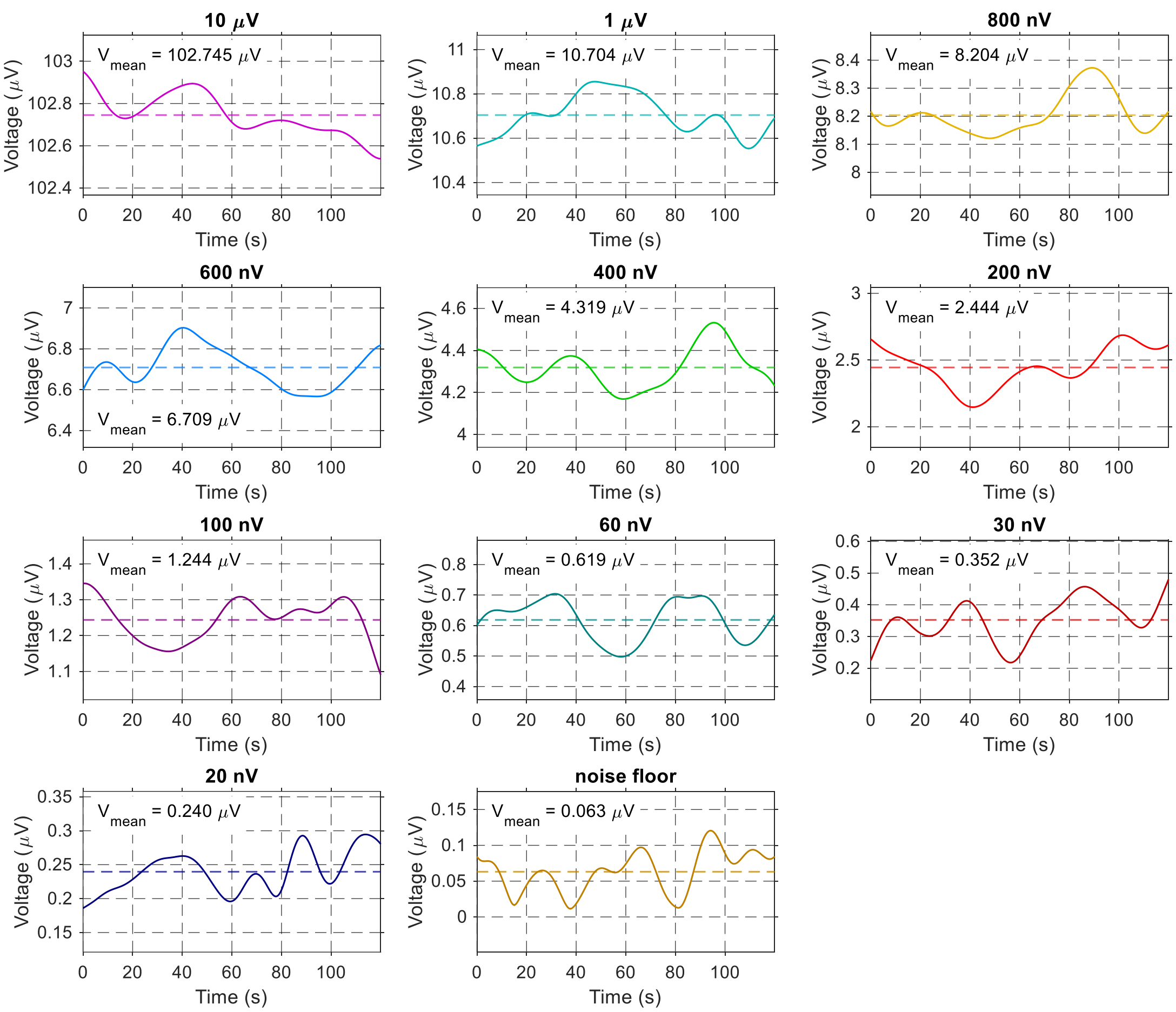

**Extended Data Fig. 9** Full raw measurement data for the 1-MHz extremely weak timing signal; solid lines represent experimental traces and dashed lines denote the averaged values.

**Extended Data Table 1** Performance comparison with state-of-the-art timing detection methods

| Works | Optical power (mW) per PD[*] | Real-time timing noise floor ($s^2$/Hz) | INTJ @ [1 kHz 1MHz] (as) | $\mathrm{LUI_T}$ (fs) | Dynamic range (dB)[***] |
|---|---|---|---|---|---|
| Ref. 3 | >80 | $1.27 \times 10^{-42}$ | 9 | 100 | 80.9 |
| Ref. 5 | > 150 | $4.3 \times 10^{-43}$ | 100 | 200 | 66 |
| Ref. 6 | 14 | $2.8 \times 10^{-43}$ | >200 | 300 | <63.5 |
| Ref. 8 | > 300 | $2 \times 10^{-42}$[**] | 1000 | 500 | 54.0 |
| Ref. 9 | 2 | $1 \times 10^{-40}$ | 16 | 500 | 89.9 |
| Our work | 1.2 | $1.61 \times 10^{-46}$ | 0.1 | 6150 | 155.8 |

[*]Total fundamental power before PD for nonlinear timing detectors (refs. 3, 5, 8); total input power of PD for others.

[**]The value of $1.5 \times 10^{-44}$ $s^2$/Hz reported in the paper is the result of 70,000 cross-correlations accumulated over 4 hours.

[***]Calculated based on the data from the $\mathrm{LUI_T}$ and INTJ columns.

## Supplementary Information

# Integrated yoctosecond-precision timing detector

**Jie Yang,[1†] Tong Wang,[1†] Yulin Shen,[1] Dehui Pan,[1] Zhichao Chen,[2] Ming-Yang Zheng,[3*] Bin Wang,[3] Shaobo Fang,[4] Yi Zhang,[1] Ke Zhang,[1] Jiahui Yao,[1] Minghua Chen,[5*] Guorong Wu,[6*] Xueming Yang[2,6], and Ming Xin[1*]**

[1]*School of Electrical and Information Engineering, Tianjin University, Tianjin, 300072, China*
[2]*Dalian Institute of Chemical Physics, Dalian, 116023, China*
[3]*Jinan Institute of Quantum Technology, Jinan 250102, China*
[4]*Institute of Physics, Chinese Academy of Sciences, Beijing, 100190, China*
[5]*Department of Electronic Engineering, Tsinghua University, Beijing, 100084, China*
[6]*Institute of Advanced Light Source Facilities, Shenzhen 518107, China*

[†]These authors contributed equally to this work.
[*]Email: zhengmingyang@jiqt.org, chenmh@tsinghua.edu.cn, wugr@mail.iasf.ac.cn, xinm@tju.edu.cn

## Contents

# I. Theoretical model

## Principle of DEST

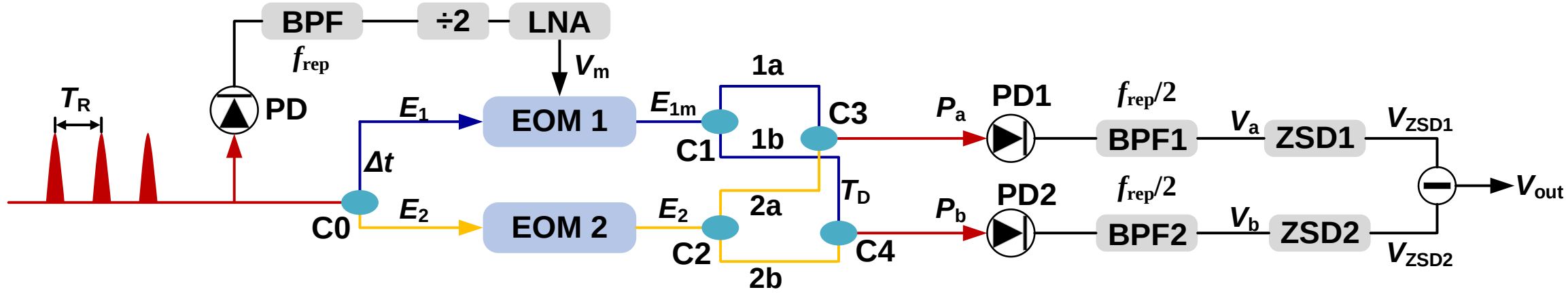

**Figure S1.** Detailed principle of DEST. EOM, electro-optic modulator; PD, photodetector; BPF, band pass filter; ÷2, frequency divider; LNA, low noise amplifier; C*i* (*i*=0, …4), 1×2 50:50 coupler; ZSD, zero-bias Schottky diode.

The detailed working principle of dual electro-optic sub-cycle timing detector (DEST) is given in Fig. S1. Without loss of generality, transmission losses of all devices are neglected. The optical pulse train (repetition rate $f_{\text{rep}}$) generated by a mode-locked laser is split into three paths. In the first path, the pulses undergo photodetection, followed by bandpass filtering, frequency division by a factor of two, and subsequent amplification—yielding a square wave voltage signal that drives EOM 1:

$$V(t)=\frac{1}{2}\sum_{k=-\infty}^{+\infty}u(t-kT_{\text{R}})(1+e^{-jk\pi})V_{\pi} \tag{S1}$$

where $T_{\text{R}} = 1/f_{\text{rep}}$ is the period of the pulse train, $V_{\pi}$ is the half-wave voltage of the EOM, and

$$u(t)=\begin{cases}1, & 0\le t<T_{\text{R}}\\ 0, & t<0 \text{ or } t>T_{\text{R}}\end{cases} \tag{S2}$$

The other two paths enter EOM1 and EOM2, respectively, with corresponding electric field strengths:

$$E_1(t)=\sum_{k=-\infty}^{+\infty}A(t-\Delta t-kT_{\text{R}})e^{-j\omega_0(t-\Delta t)} \tag{S3}$$

$$E_2(t)=\sum_{k=-\infty}^{+\infty}A(t-kT_{\text{R}})e^{-j\omega_0 t} \tag{S4}$$

where $A(t)$ is the pulse envelope, $\omega_0$ is the angular frequency of the optical carrier, and $\Delta t$ is the relative timing jitter between the two pulse trains.

Based on Eq. (S1-S3), the modulated optical pulse train after EOM 1 is:

$$E_{1\mathrm{m}}(t)=\sum_{k=-\infty}^{+\infty}(-1)^k A(t-\Delta t-kT_{\mathrm{R}})e^{-j\omega_0(t-\Delta t)} \tag{S5}$$

Through couplers C1 and C2, the powers of $E_{1\mathrm{m}}$ and $E_2$ are equally split into paths 1a, 1b and 2a, 2b, respectively. The powers in paths 1a and 2a are then combined by coupler C3. Finally, the optical power before PD1 can be written as:

$$P_{\mathrm{a}}(t)\propto\frac{1}{4}\left|E_{1\mathrm{m}}(t)+E_2(t)\right|^2 \tag{S6}$$

Substituting Eq. (S4) and (S5) into Eq. (S6), we obtain

$$P_{\mathrm{a}}(t)\propto\frac{1}{4}\sum_{k=-\infty}^{+\infty}\left[A(t-kT_{\mathrm{R}})^2+A(t-\Delta t-kT_{\mathrm{R}})^2\right]+\frac{1}{2}\cos(\omega_0\Delta t)\sum_{k=-\infty}^{+\infty}(-1)^k A(t-kT_{\mathrm{R}})A(t-\Delta t-kT_{\mathrm{R}}) \tag{S7}$$

Only the second term on the right-hand side of Eq. (S7) contributes to odd harmonics of $f_{\mathrm{rep}}/2$. Therefore, after photodetection and bandpass filtering with BPF1 centered at $f_{\mathrm{rep}}/2$, the output voltage can be written as:

$$V_{\mathrm{a}}(t)\propto\frac{1}{2}\cos(\omega_0\Delta t)\cos(\pi f_{\mathrm{rep}}t)\int_{-\infty}^{+\infty}A(t)A(t-\Delta t)dt \tag{S8}$$

Similarly, it can be obtained:

$$V_{\mathrm{b}}(t)\propto\frac{1}{2}\cos\left[\omega_0(\Delta t+T_{\mathrm{D}})\right]\cos(\pi f_{\mathrm{rep}}t)\int_{-\infty}^{+\infty}A(t)A(t-\Delta t-T_{\mathrm{D}})dt \tag{S9}$$

Suppose $A(t)$ is a hyperbolic secant function [Ref. S1]:

$$A(t)=A_0\,\mathrm{sech}\left(\frac{t}{\tau}\right) \tag{S10}$$

where $A_0$ is the envelope amplitude and $\tau$ is the pulse width parameter. Then

$$\int_{-\infty}^{+\infty}A(t)A(t-\Delta t)dt=A_0^2\frac{4\Delta t e^{\Delta t/\tau}}{e^{2\Delta t/\tau}-1} \tag{S11}$$

Substituting Eq. (S11) into Eq. (S8) and (S9), we have

$$V_{\mathrm{a}}(t)=V_0\cos(\omega_0\Delta t)\cos(\pi f_{\mathrm{rep}}t)F\left(\frac{\Delta t}{\tau}\right) \tag{S12}$$

$$V_{\mathrm{b}}(t)=V_0\cos\left[\omega_0(\Delta t+T_{\mathrm{D}})\right]\cos(\pi f_{\mathrm{rep}}t)F\left(\frac{\Delta t+T_{\mathrm{D}}}{\tau}\right) \tag{S13}$$

where $V_0$ is a constant voltage related to the incident light power, the responsivity of the photodetector, and the transimpedance gain, and $F(x)$ is a dimensionless function with the following form:

$$F(x)=\frac{2xe^{x}}{e^{2x}-1} \tag{S14}$$

Define $H_{\mathrm{ZSD}}(V)$ as the transfer function of the ZSD, then the output voltage of the two ZSDs are

$$V_{\mathrm{ZSD1}}(\Delta t)=H_{\mathrm{ZSD}}\left[V_0F\left(\frac{\Delta t}{\tau}\right)\left|\cos(\omega_0\Delta t)\right|\right] \tag{S15}$$

$$V_{\mathrm{ZSD2}}(\Delta t)=H_{\mathrm{ZSD}}\left[V_0F\left(\frac{\Delta t+T_D}{\tau}\right)\left|\cos\left(\omega_0(\Delta t+T_D)\right)\right|\right] \tag{S16}$$

And the final differential output voltage is

$$V_{\mathrm{out}}(\Delta t)=V_{\mathrm{ZSD2}}(\Delta t)-V_{\mathrm{ZSD1}}(\Delta t) \tag{S17}$$

## Transfer function of ZSD

All ZSDs used in the experiment are homemade. Each ZSD consists of a Schottky diode followed by a parallel resistor-capacitance (RC) network at the output port. The Schottky diode has a typical total capacitance of 0.3 pF, a forward voltage drops of 60–120 mV at 0.1 mA and 135–240 mV at 1 mA, and a typical video resistance of 5000 Ω. The parallel RC network has a theoretical 3-dB bandwidth of approximately 25.5 MHz; it filters and smooths the high-frequency components after diode rectification, yielding an output voltage that depends on the amplitude of the input RF signal.

**Table S1.** Main parameters used in LTspice simulation

| Parameter | Description | Value | Unit |
|---|---|---|---|
| $I_S$ | saturation current | $5\times 10^{-6}$ | A |
| $R_S$ | ohmic resistance | 20 | Ω |
| $\eta$ | emission coefficient | 1.05 | - |
| $C_{JO}$ | zero-bias junction capacitance | 0.14 | pF |
| M | grading coefficient | 0.40 | - |
| $E_G$ | band-gap energy | 0.69 | eV |
| $B_V$ | reverse breakdown voltage | 2 | V |
| $C_L$ | load capacitance | 16 | pF |
| $R_L$ | load resistance | 390 | Ω |

Using the simulation parameters listed in Table S1, the ZSD performance was simulated with LTspice, yielding the input-output relationship shown in Fig. S2. Treating the output voltage as a polynomial function of the input, below 0.2 V the output is dominated by quadratic and higher-order terms, while above 0.2 V it gradually becomes linear. Consequently, a single polynomial cannot accurately describe the ZSD behavior. We therefore adopt an interpolation strategy: based on the discrete data in Fig. S2, spline interpolation provides the output voltage for any input voltage, thereby determining the transfer function $H_{ZSD}(V)$ of the ZSD.

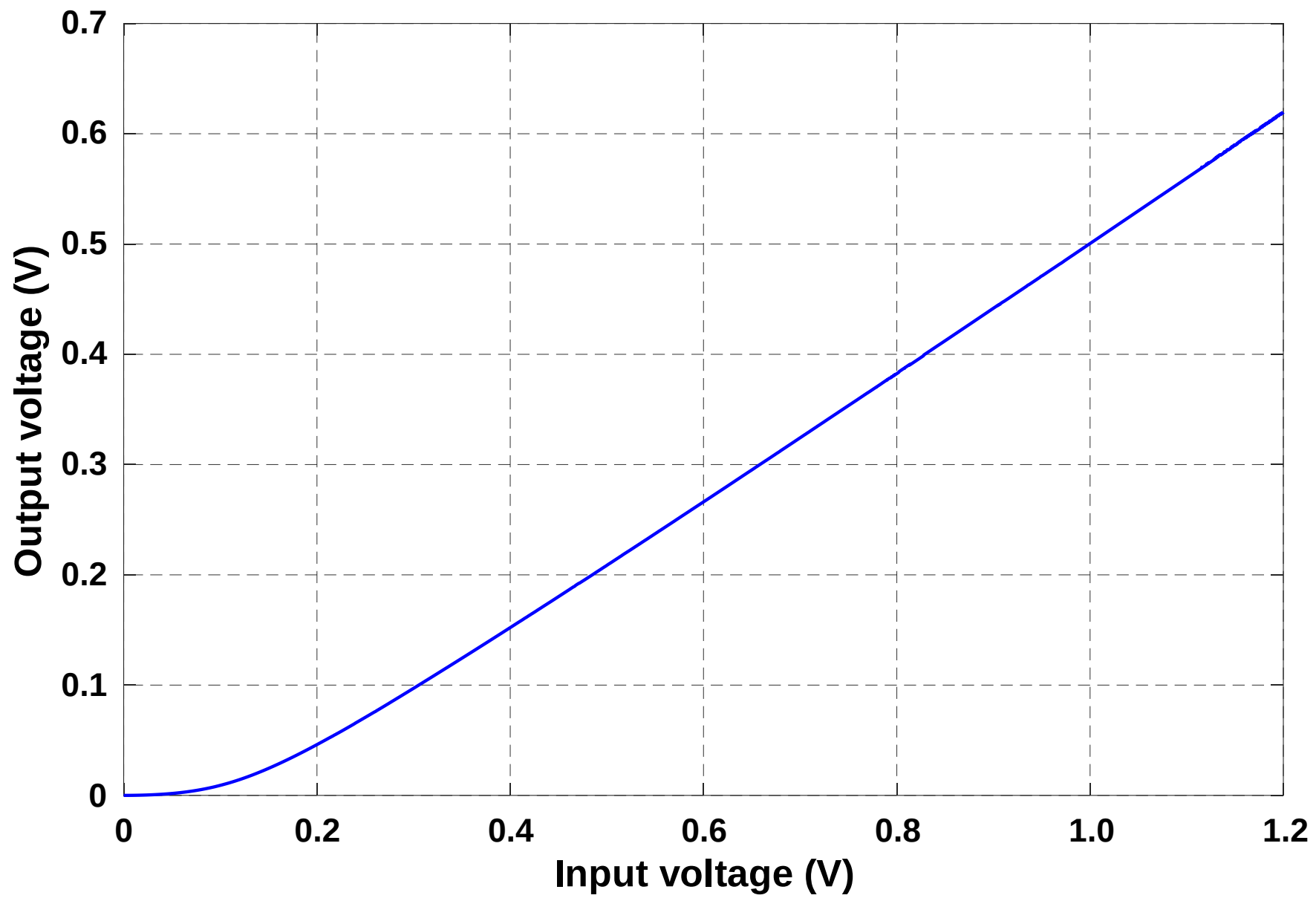

**Figure S2.** The relation between the input and output voltages of the ZSD.

## Timing characteristic curve

Using the interpolated transfer function $H_{ZSD}$ together with Eq. (S15) and (S16), the relationship between the timing jitter $\Delta t$ and the voltages $V_{ZSD1}$ and $V_{ZSD2}$ can be computed. To facilitate the subsequent analysis of the unambiguous interval, $V_{ZSD1}$ and $V_{ZSD2}$ are normalized as follows:

$$V_1(\Delta t) = \frac{H_{ZSD}\left[V_0 F\left(\frac{\Delta t}{\tau}\right)\left|\cos(\omega_0 \Delta t)\right|\right]}{H_{ZSD}(V_0)} \tag{S18}$$

$$V_2(\Delta t) = \frac{H_{ZSD}\left[V_0 F\left(\frac{\Delta t + T_D}{\tau}\right)\left|\cos\left(\omega_0(\Delta t + T_D)\right)\right|\right]}{H_{ZSD}(V_0)} \tag{S19}$$

where $V_0$ is taken from experimental measurement as 1V. This normalization is convenient because the quantization resolution of an analog-to-digital converter (ADC) depends only on its bit number; once normalized, the resolution is expressed simply as $2^{-n}$ without needing to consider the actual input voltage range.

Fig. S3 shows $V_1$ and $V_2$ as functions of $\Delta t$. Both voltages are periodic functions modulated by an envelope; they drop to zero every $T_c/2$, where $T_c=2\pi/\omega_0$ is the optical carrier period. Using $T_c$, the relative delay $T_D$ between the two curves can be redefined as $T_D = N \times T_c / 2 + t_d$, where $N$ is a nonnegative integer and $0 \leq t_d < T_c/2$.

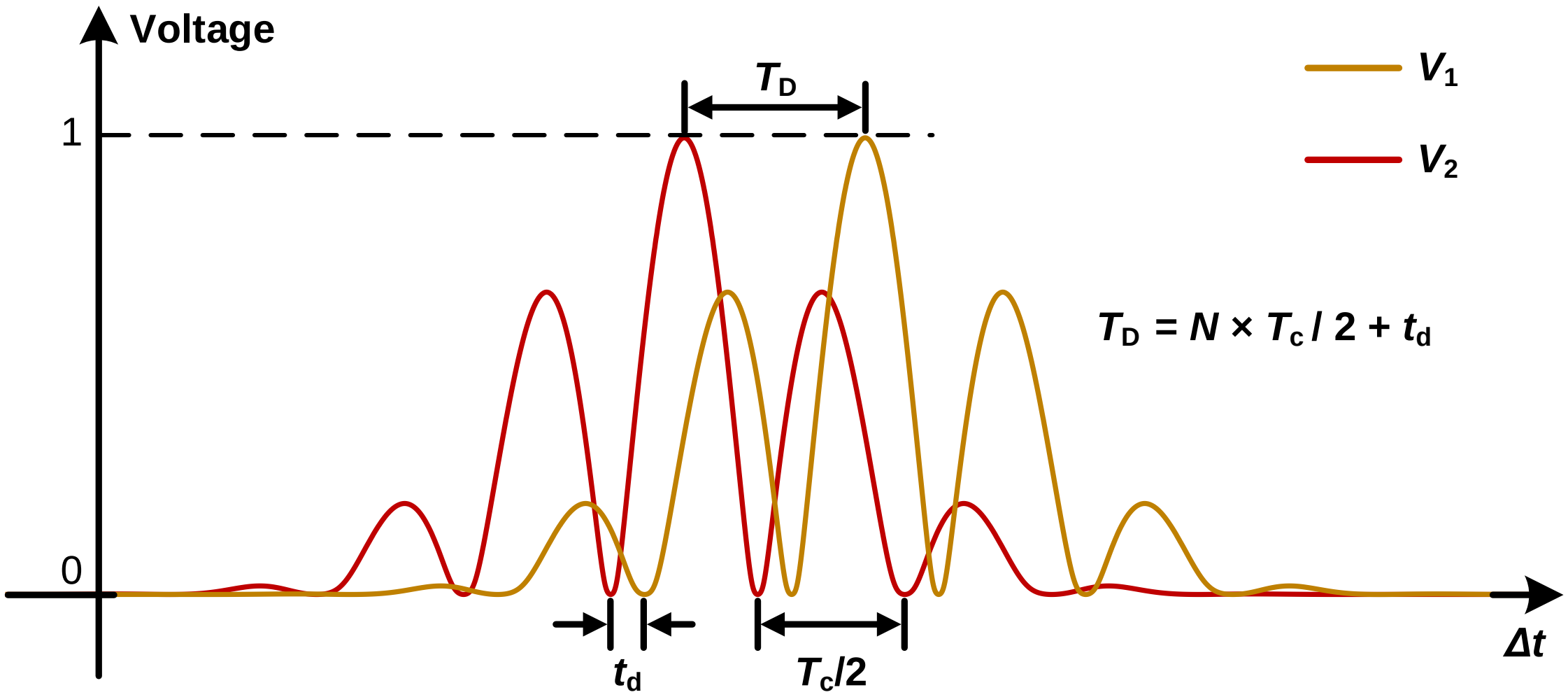

**Figure S3.** The relation between the output voltages of the two ZSDs and $\Delta t$. $T_c$=5.2 fs (corresponding wavelength $\lambda_0 \approx$ 1558.9 nm), to reveal the details of the curves, $\tau$ is set to 1.7 fs.

Define the normalized differential output as $V_o=V_2-V_1$. Fig. S4 shows $V_o$ as a function of $\Delta t$, i.e., the normalized timing characteristic curve. This curve consists of several sub cycles. Within each sub cycle the curve is monotonic (strictly increasing or decreasing). Every sub cycle has a zero-crossing. We define $\Delta t_{TS}$ as the zero-crossing that possesses the largest absolute slope among all sub cycles, and denote that maximum absolute slope as the normalized timing sensitivity $TS_N$:

$$TS_N = \max_{\Delta t \in \mathbb{Z}} \left| \frac{dV_o(\Delta t)}{d\Delta t} \right|, \quad \mathbb{Z} = \{\Delta t | V_o(\Delta t) = 0\} \tag{S20}$$

$$\Delta t_{TS} = \arg\max_{\Delta t \in \mathbb{Z}} \left| \frac{dV_o(\Delta t)}{d\Delta t} \right|, \quad \mathbb{Z} = \{\Delta t | V_o(\Delta t) = 0\} \tag{S21}$$

For the $n$-th sub cycle we introduce a feature vector $\mathbf{C}_n = [C_{n,1}, C_{n,2}, C_{n,3}, C_{n,4}]$, where the four components are the maximum voltage, the minimum voltage, the average voltage, and the polarity (+1 for a monotonically increasing sub cycle, –1 for a decreasing one), respectively. To determine whether any two sub cycles $n$ and $k$ are distinguishable, we construct a similarity matrix $\mathbf{M}$:

$$M(n,k) = \begin{cases} \sum_{q=1}^{3} \Theta\left(\bar{V}_{th} - |C_{n,q} - C_{k,q}|\right) + \frac{1}{2}|C_{n,4} - C_{k,4}|, & n \neq k \\ 0, & n = k \end{cases} \tag{S22}$$

with $\Theta(x)$=0 for $x$<0 and $\Theta(x)$=1 for $x \geq 0$; $\bar{V}_{th}$ is a preset normalized voltage threshold (normalized resolution). Two sub cycles $n$ and $k$ are said to be distinguishable if and only if $M(n, k)$<4.

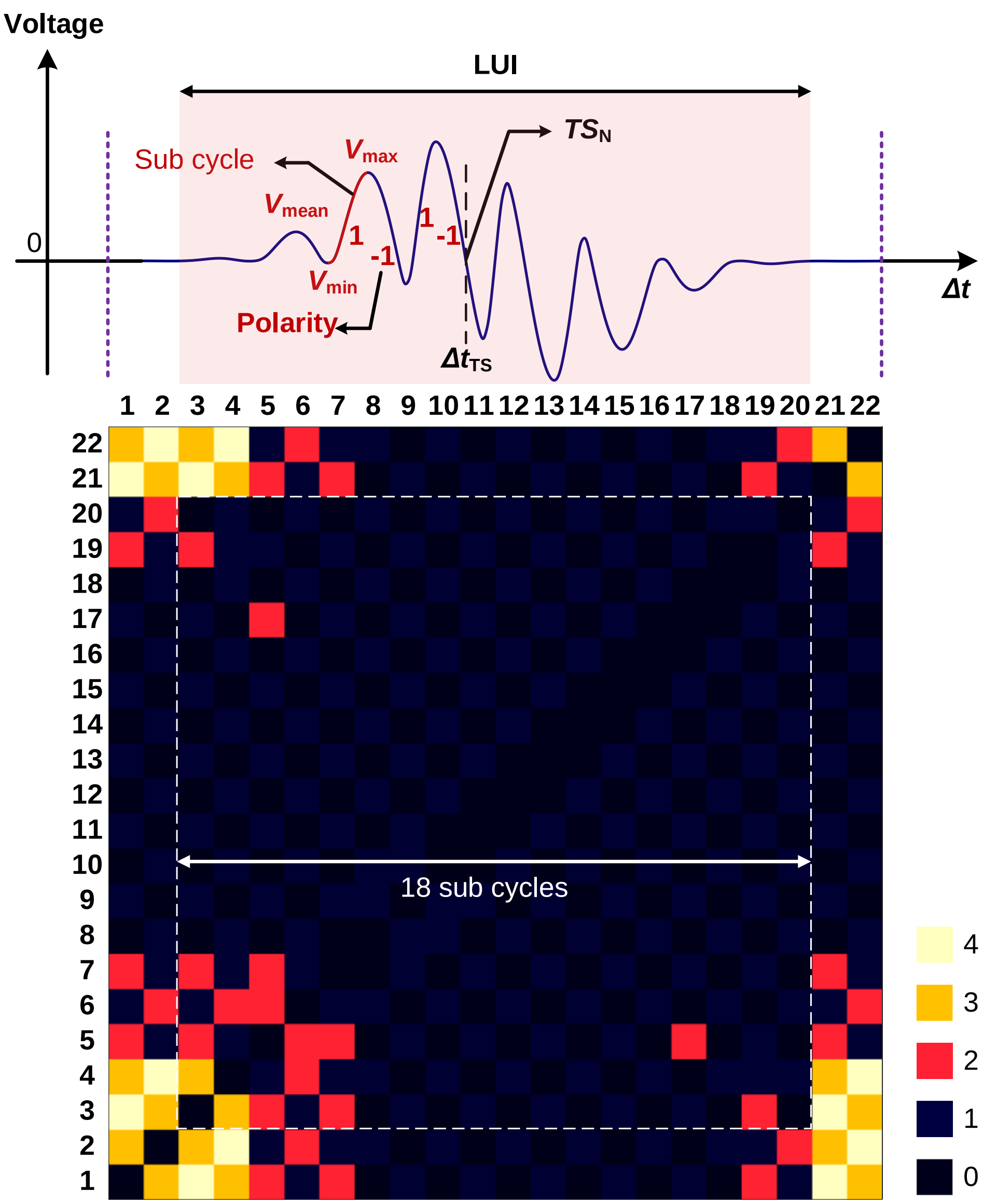

**Figure S4.** A normalized timing characterization curve and the corresponding similarity matrix heatmap. $T_c = 5.2$ fs, $\tau = 1.7$ fs, $\bar{V}_{th} = 2^{-8}$.

An unambiguous interval is defined as a contiguous range of $\Delta t$ such that any two distinct sub cycles within that range are distinguishable. In other words, when $\Delta t$ falls inside an unambiguous interval, measuring the four characteristic parameters of the current sub cycle uniquely identifies the absolute position of $\Delta t$ along the timing curve. The timing curve may contain multiple non-overlapping unambiguous intervals. Among them, the Longest Unambiguous Interval (LUI) is the one with the greatest length (i.e., covering the largest number of consecutive sub cycles).

Due to the decaying amplitude of sub cycles away from the envelope center, when the absolute values of all three voltage features (maximum, minimum, average) fall below $\bar{V}_{th}$, no unambiguous interval of length greater than two can exist. To locate the LUI, the following procedure can be used:

1. Select a sufficiently large time range covering all sub cycles with appreciable amplitude.
2. Compute the similarity matrix **M** for all sub cycles in this range.
3. Identify all contiguous blocks of sub cycle indices (diagonal submatrices) where every element is <4. The longest such block corresponds to the LUI.

As an example, in Fig. S4, we first select 22 sub cycles of the timing characteristic curve lying between the two purple dashed lines. The corresponding similarity matrix **M** is displayed as a heatmap in the lower part of Fig. S4. Within this matrix, the largest diagonal submatrix whose every entry is less than 4 is indicated by the white dashed square; it measures 18 ×18. Therefore, the LUI for this timing characteristic curve is $18 \times T_c / 4 =$ 23.4 fs.

DEST is typically operated at $\Delta t_{TS}$ where the timing sensitivity is highest. When the operating point deviates from $\Delta t_{TS}$, a scan voltage is applied to EOM2 while the DEST output is acquired by a data acquisition card. The sub cycle voltage features ($V_{max}$, $V_{mean}$, $V_{min}$, and polarity) are then extracted, and the similarity matrix is used to determine the absolute position of the operating point on the timing curve, so as to enable feedback to restore the operating point.

To ensure robust operation under external perturbations, it is desirable to choose parameter settings such that $\Delta t_{TS}$ falls well inside the LUI, leaving a large margin on both sides before the operating point exits the unambiguous range. This margin is quantified by $LUI_T$, defined as twice the shorter distance from $\Delta t_{TS}$ to the LUI boundaries (zero if $\Delta t_{TS}$ lies outside). A larger $LUI_T$ thus provides greater tolerance against disturbances while still allowing the operating point to be recovered via sub-cycle feature identification.

## Sampling points requirements

When the DEST output (with a scan voltage applied to EOM2) is acquired to identify the current sub cycle, the number of sampling points per sub cycle should be as small as possible to minimize the decision time. However, if too few points are acquired, $V_{max}$ and $V_{min}$ cannot be accurately resolved, and the measured values become dependent on the relative phase between the sampling clock and the scan voltage, potentially leading to incorrect sub cycle identification.

Fig. S5 shows the minimum number of sampling points $N_s$ required per sub cycle to achieve correct identification for all sub cycles within the $LUI_T$, regardless of the relative phase, for different pulse widths. For narrow pulses, the rapid variation of the envelope demands more sampling points to reduce the error in the extracted $V_{max}$ and $V_{min}$. As the pulse width increases and the envelope becomes flatter, the required number of sampling points decreases accordingly.

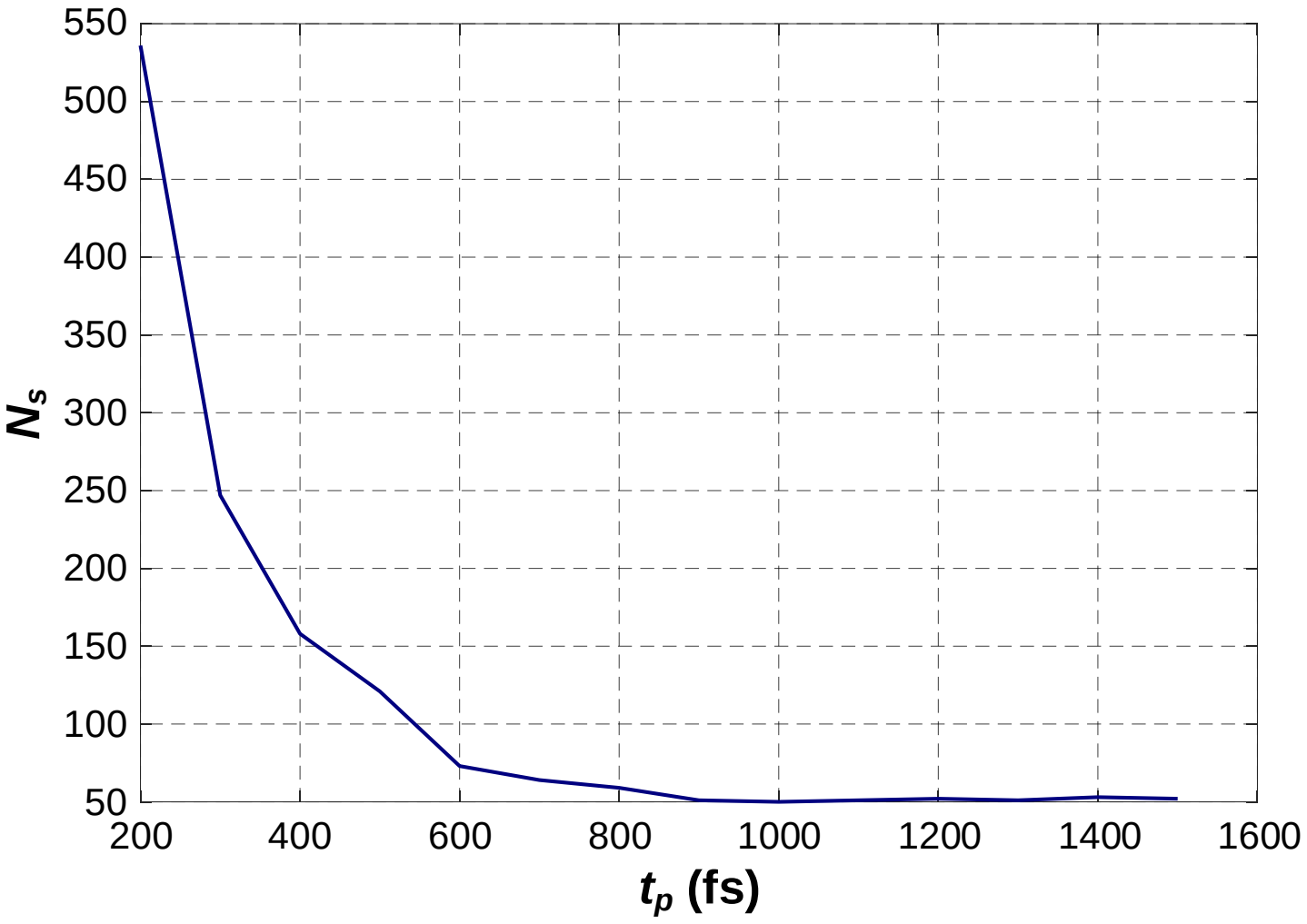

**Figure S5.** Minimum number of sampling points per sub cycle versus pulse width $t_{\mathrm{p}}$. $T_{\mathrm{c}}$ = 5.2 fs ($\lambda_0$ ≈ 1558.9 nm), $N$ = 202 $t_{\mathrm{d}}$ = 0.523 fs.

## Noise floor analysis

The timing detection noise floor of DEST is ultimately limited by four noise sources: shot noise from the optical pulses at the photodetector input, the detector's electronic noise, and the Schottky diode's shot noise and thermal noise. The calculation method for each is detailed below.

In Fig. S1, the average optical power $P_{\mathrm{in}}$ entering PD1 is provided solely by the first term on the right-hand side of Eq. (S7). The corresponding shot-noise voltage spectral density (in $\mathrm{V^2/Hz}$) is

$$v_{\mathrm{shot}}^2 = 2eR_{\mathrm{p}}P_{\mathrm{in}}G^2 \tag{S23}$$

where $e$ is the elementary charge, $R_{\mathrm{p}}$ is the photodetector responsivity, and $G$ is the transimpedance gain. With the photodetector's equivalent noise power NEP (in W/√Hz), the electronic noise voltage spectral density at the PD output is

$$v_{\mathrm{e}}^2 = (\mathrm{NEP} \times R_{\mathrm{p}} \times G)^2 \tag{S24}$$

From Eq. (S12), when $\Delta t$=0, we have $V_{\mathrm{a}}(t) = V_0\cos(\pi f_{\mathrm{rep}}t)$, where the amplitude $V_0$ arises from the second term of Eq. (S7) and corresponds to the same optical power $P_{\mathrm{in}}$. Thus,

$$V_0 = R_{\mathrm{p}}P_{\mathrm{in}}G \tag{S25}$$

Substituting Eq. (S25) into Eq. (S23) and (S24) yields

$$v_{\mathrm{shot}}^2 = \frac{2eV_0^2}{R_{\mathrm{p}}P_{\mathrm{in}}} \tag{S26}$$

$$v_{\mathrm{e}}^{2}=\left(\frac{V_{0}\times\mathrm{NEP}}{P_{\mathrm{in}}}\right)^{2} \tag{S27}$$

Since $V_{\mathrm{out}}(\Delta t_{\mathrm{TS}}) = 0$, we can write $V_{\mathrm{ZSD1}}(\Delta t_{\mathrm{TS}}) = V_{\mathrm{ZSD2}}(\Delta t_{\mathrm{TS}}) = H_{\mathrm{ZSD}}(V_{\mathrm{W}})$, where $V_{\mathrm{W}}$ is the amplitude of the RF signals $V_{\mathrm{a}}(t)$ and $V_{\mathrm{b}}(t)$ at which $V_{\mathrm{out}}$ operates at the maximum timing sensitivity $TS_{\mathrm{N}}$. When a small noise voltage $\Delta V$ is added to $V_{\mathrm{W}}$, First-order Taylor expansion yields

$$H_{\mathrm{ZSD}}(V_{\mathrm{W}}+\Delta V)\approx H_{\mathrm{ZSD}}(V_{\mathrm{W}})+\left.\frac{dH_{\mathrm{ZSD}}}{dV}\right|_{V=V_{\mathrm{W}}}\times\Delta V \tag{S28}$$

Consequently, after the ZSD, both the shot noise and the PD electronic noise are amplified by the factor $H_{\mathrm{ZSD}}'(V_{\mathrm{W}})$:

$$v_{\mathrm{shot,amp}}^{2}=\left[H_{\mathrm{ZSD}}'(V_{\mathrm{W}})\right]^{2}\frac{2eV_{0}^{2}}{R_{\mathrm{p}}P_{\mathrm{in}}} \tag{S29}$$

$$v_{\mathrm{e,amp}}^{2}=\left(H_{\mathrm{ZSD}}'(V_{\mathrm{W}})\times\frac{V_{0}\times\mathrm{NEP}}{P_{\mathrm{in}}}\right)^{2} \tag{S30}$$

The thermal noise voltage spectral density (in $\mathrm{V}^2/\mathrm{Hz}$) of the ZSD is given by the Johnson–Nyquist formula:

$$v_{\mathrm{ZSD,n}}^{2}=4k_{\mathrm{B}}TR_{\mathrm{S}} \tag{S31}$$

where $k_{\mathrm{B}}$ is Boltzmann's constant, and $T$ is absolute temperature.

The ZSD's current–voltage characteristic follows the Shockley diode equation:

$$I=I_{\mathrm{S}}\left[\exp\left(\frac{V-IR_{\mathrm{S}}}{\eta k_{\mathrm{B}}T/e}\right)-1\right] \tag{S32}$$

For an input voltage of $V_{\mathrm{W}}\cos(\pi f_{\mathrm{rep}}t)$, the instantaneous current $i(t)$ satisfies:

$$i(t)=I_{\mathrm{S}}\left[\exp\left(\frac{V_{\mathrm{W}}\cos(\pi f_{\mathrm{rep}}t)-i(t)R_{\mathrm{S}}}{\eta k_{\mathrm{B}}T/e}\right)-1\right] \tag{S33}$$

The output DC current is obtained by integrating over the phase space:

$$I_{\mathrm{DC}}=\frac{1}{2\pi}\int_{0}^{2\pi}i(\theta)d\theta \tag{S34}$$

For each phase angle $\theta$, $i(\theta)$ is solved numerically from the transcendental equation:

$$i(\theta)=I_{\mathrm{S}}\left[\exp\left(\frac{V_{\mathrm{W}}\cos(\theta)-i(\theta)R_{\mathrm{S}}}{\eta k_{\mathrm{B}}T/e}\right)-1\right] \tag{S35}$$

With $I_{\mathrm{DC}}$ known, the shot-noise voltage spectral density at the ZSD output (in $\mathrm{V}^2/\mathrm{Hz}$) is:

$$v_{\text{ZSD,shot}}^{2} = 2eI_{\text{DC}}R_{\text{L}}^{2} \tag{S36}$$

Thus, the total noise voltage spectral density at the differential output port is:

$$v_{\text{all}}^{2} = 2\times\left(v_{\text{shot,amp}}^{2} + v_{\text{e,amp}}^{2} + v_{\text{ZSD,n}}^{2} + v_{\text{ZSD,shot}}^{2}\right) \tag{S37}$$

Since the actual timing sensitivity after differential operation is $H_{\text{ZSD}}(V_0) \times TS_{\text{N}}$ (in V/s), the timing noise power spectral density (in s²/Hz) is:

$$J_{\text{all}}^{2} = \frac{2\times\left(v_{\text{shot,amp}}^{2} + v_{\text{e,amp}}^{2} + v_{\text{ZSD,n}}^{2} + v_{\text{ZSD,shot}}^{2}\right)}{\left[H_{\text{ZSD}}(V_0)TS_{\text{N}}\right]^{2}} = J_{\text{shot,amp}}^{2} + J_{\text{e,amp}}^{2} + J_{\text{ZSD,n}}^{2} + J_{\text{ZSD,shot}}^{2} \tag{S38}$$

where $J_{\text{shot,amp}}^{2}$, $J_{\text{e,amp}}^{2}$, $J_{\text{ZSD,n}}^{2}$, $J_{\text{ZSD,shot}}^{2}$ represent the contributions from the four noise sources. Let $P_i = 2P_{\text{in}}$ be the total average optical power entering the two photodetectors. For a fixed number of periods $N$, $J_{\text{all}}^{2}$ is a function of both $P_i$ and $t_{\text{d}}$. We define $TJ_{\text{all}}^{2}$ as the minimum value of $J_{\text{all}}^{2}$ achieved by varying $t_{\text{d}}$:

$$TJ_{\text{all}}^{2}(P_i) = \min_{0\le t_d < T_c/2} J_{\text{all}}^{2}(P_i, t_d) \tag{S39}$$

The corresponding optimal value $t_{\text{d}}^{*}$ that minimizes $J_{\text{all}}^{2}$ is thus a function of $P_i$:

$$t_{\text{d}}^{*}(P_i) = \arg\min_{0\le t_{\text{d}} < T_{\text{c}}/2} J_{\text{all}}^{2}(P_i, t_{\text{d}}) \tag{S40}$$

At this optimal $t_{\text{d}}^{*}$, the contributions of the four noise sources become

$$TJ_{\text{shot,amp}}^{2}(P_i) = J_{\text{shot,amp}}^{2}\left(P_i, t_{\text{d}}^{*}(P_i)\right) \tag{S41}$$

$$TJ_{\text{e,amp}}^{2}(P_i) = J_{\text{e,amp}}^{2}\left(P_i, t_{\text{d}}^{*}(P_i)\right) \tag{S42}$$

$$TJ_{\text{ZSD,n}}^{2}(P_i) = J_{\text{ZSD,n}}^{2}\left(P_i, t_{\text{d}}^{*}(P_i)\right) \tag{S43}$$

$$TJ_{\text{ZSD,shot}}^{2}(P_i) = J_{\text{ZSD,shot}}^{2}\left(P_i, t_{\text{d}}^{*}(P_i)\right) \tag{S44}$$

Fig. S6 plots these four contributions together with the total $TJ_{\text{all}}^{2}$ as functions of $P_i$ for two wavelengths. For both wavelengths, the following trends are observed: the PD electronic noise dominates below 0.2 mW; the PD shot noise prevails between 1 mW and 100 mW; and the ZSD thermal noise becomes dominant near 1 W. Taking the square root of the $TJ_{\text{all}}^{2}$ data in panels (a) and (b) yields the blue solid and dashed curves in Fig. 1f, respectively.

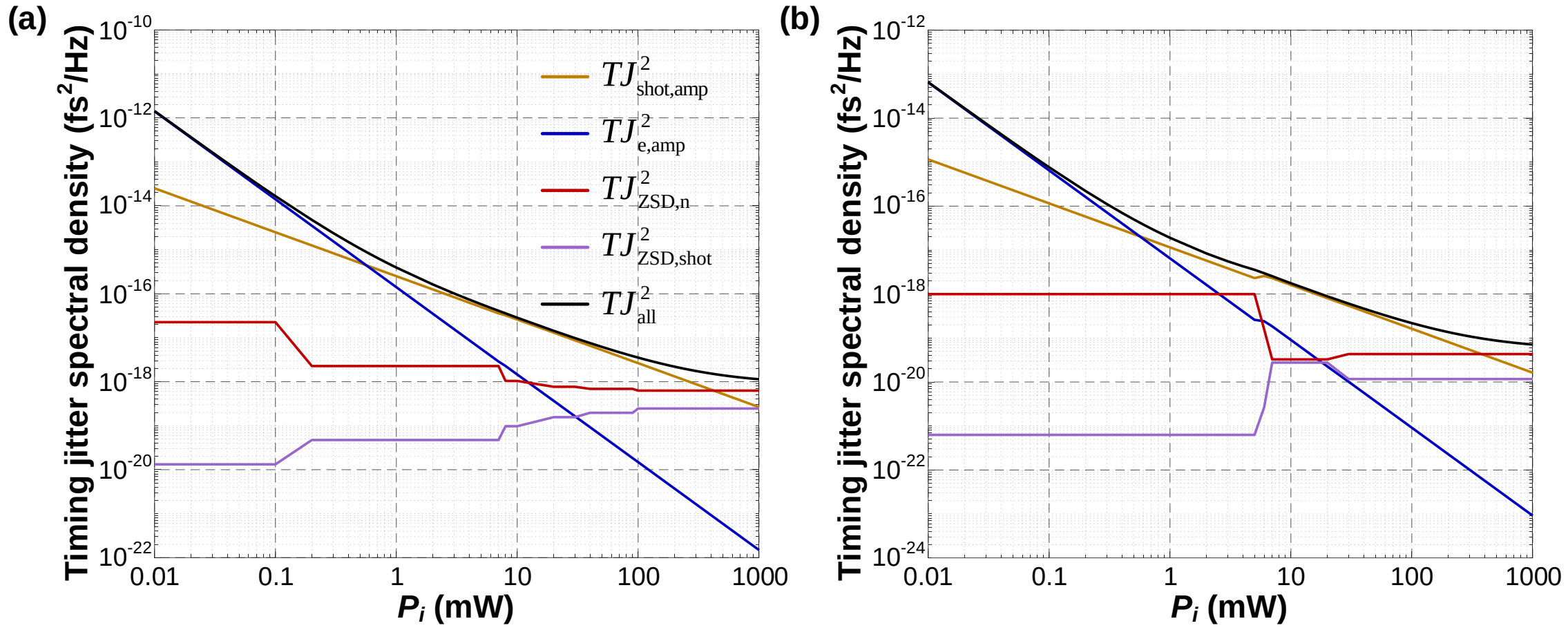

**Figure S6.** Timing jitter spectral density versus the total input optical power $P_i$ incident on the two photodetectors for (a) $T_c = 5.2$ fs ($\lambda_0 \approx 1558.9$ nm), $N = 202$ and (b) $T_c = 1.28$ fs ($\lambda_0 \approx 383.7$ nm), $N = 808$. Common parameters: pulse width $t_p = 1.2$ ps, room temperature $T = 300$ K, PD responsivity $R_p = 0.9$ A/W, and PD NEP = 10 pW/√Hz. Each panel displays the contributions of the four individual noise sources together with the total $TJ^2_{\text{all}}$.

With a measurement bandwidth BW, the dynamic range (DR, in dB) is:

$$\text{DR} = 20 \times \log_{10}\left(\frac{\text{LUI}_\text{T}}{\sqrt{TJ^2_{\text{all}} \times \text{BW}}}\right) \tag{S45}$$

where $\text{LUI}_\text{T}$ is evaluated at the same $t^*_\text{d}(P_i)$ used for achieving $TJ^2_{\text{all}}$.

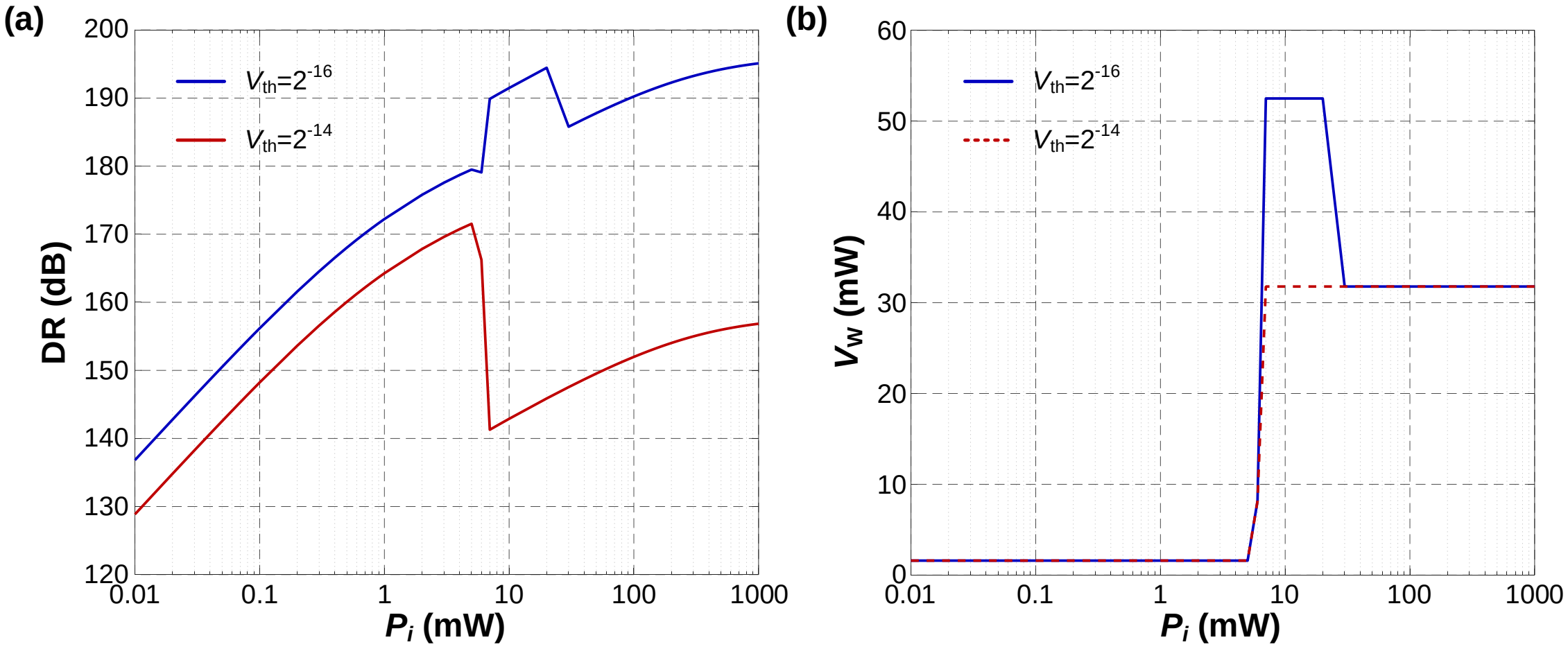

**Figure S7.** (a) Dynamic range (DR) and (b) operating voltage $V_W$ versus the total input optical power $P_i$ for two threshold voltages: $\bar{V}_{th} = 2^{-14}$ and $\bar{V}_{th} = 2^{-16}$. Common parameters: $t_p = 1.2$ ps, $T_c = 1.28$ fs ($\lambda_0 \approx 383.7$ nm), $N = 808$, BW=1 MHz.

For $\lambda_0 = 1558.9$ nm, $N = 202$ and $\bar{V}_{th} = 2^{-14}$, the DR as a function of optical power $P_i$ is shown by the red solid curve in Fig. 1f; for $\lambda_0 = 383.7$ nm, $N = 808$ and $\bar{V}_{th} = 2^{-16}$, it is shown by the red dashed curve. Fig. S7a further compares the DR for $\lambda_0 = 383.7$ nm at two threshold voltages: $\bar{V}_{th} = 2^{-14}$ and $\bar{V}_{th} = 2^{-16}$. As shown in Fig. S7b, the discontinuities in the DR curves arise from abrupt changes in the operating voltage $V_W$: at $P_i = 5$ mW for $\bar{V}_{th} = 2^{-14}$ and $P_i = 5$ mW and 20 mW for $\bar{V}_{th} = 2^{-16}$.

## II. Design of integrated components

This section details the design strategies for the integrated DEST, aimed at achieving both a low detection noise floor and a wide LUI.

### Vertical architecture design

The integrated DEST is fabricated on an X-cut lithium-niobate-on-insulator (LNOI) wafer, which consists of a 525-µm-thick silicon substrate, a 2-µm-thick $SiO_2$ layer, and a 500-nm-thick lithium niobate (LN) film. To protect the LN waveguides from contamination, a $SiO_2$ cladding layer is deposited on top of the LN film by plasma-enhanced chemical vapor deposition (PECVD), significantly improving the device robustness against environmental disturbances during operation.

With the cladding layer introduced, two schemes are available for electrode configuration. In Scheme A, the electrode is deposited directly on the LN film and connected to bonding pads on the cladding surface through vias (Fig. S8a). In Scheme B, the electrode is deposited directly on the cladding (Fig. S8b). Scheme A requires three rounds of photolithography and reactive-ion etching (RIE) to define the electrodes, vias, and pads, whereas Scheme B requires only one round for electrode patterning, substantially reducing cost and cumulative fabrication errors. Therefore, we adopt Scheme B in our final design to minimize processing inaccuracies.

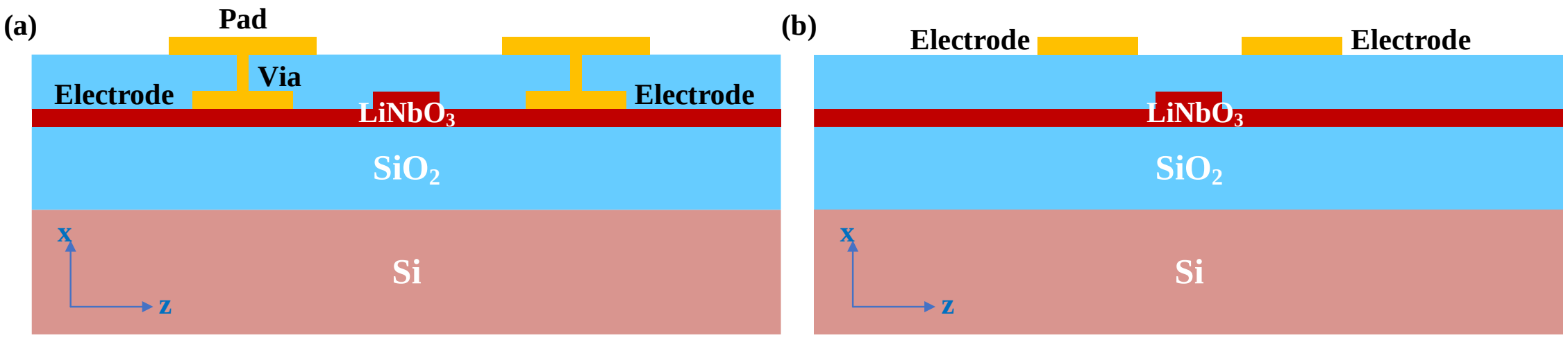

**Figure S8.** Electrode placement schemes: (a) on the LN film; (b) on the cladding.

### Electro-optic modulator

The electro-optic modulator adopts an LN ridge waveguide structure. The ridge geometry provides both lateral and vertical optical confinement while retaining a slab layer of finite thickness, which reduces etching loss and improves both transmission stability and detection sensitivity of the timing detector. Fig. S9a shows the RF electric field distribution when a voltage is applied to the gold electrodes; the field is primarily concentrated in the gap between the two electrodes and within the LN waveguide region, enhancing the electro-optic interaction. Fig. S9b displays the quasi-TE mode distribution at 1550 nm, where the optical power is well confined within the LN ridge and exhibits good spatial overlap with the electric-field region. Consequently, this structure enables efficient on-chip phase modulation, providing the basis for the modulation of the $f_{rep}/2$ signal and the scanning voltage required in the integrated DEST.

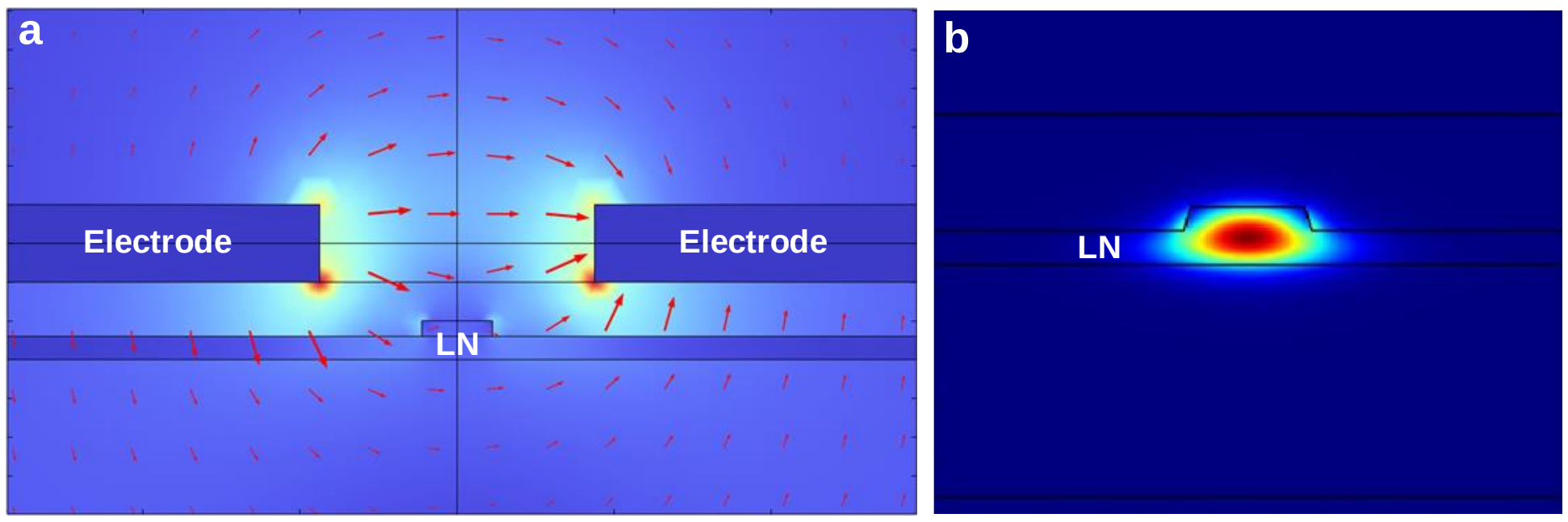

**Figure S9.** (a) RF electric-field distribution and (b) optical-field distribution around the LN ridge waveguide.

The design parameters of the ridge waveguide are given in Fig. S10a. Simulation results show that the half-wave voltage $V_\pi$ is relatively insensitive to waveguide width variations, , as the optical-RF field overlap remains nearly constant as long as the quasi-TE mode is stably supported. In contrast, $V_\pi$ depends more strongly on the LN film thickness $h_{LN}$ and etching depth $h_{wg}$, since both affect the vertical mode profile and its overlap with the RF field.

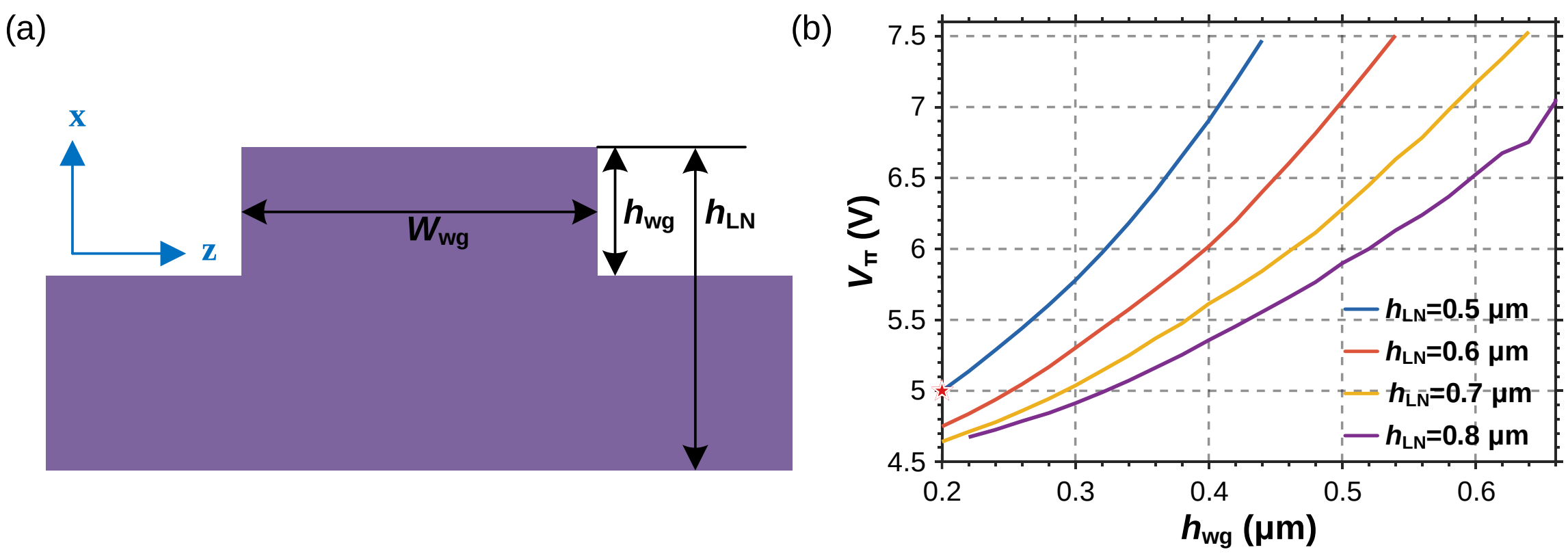

**Figure S10.** (a) Parameter definitions of the LN ridge waveguide; (b) $V_\pi$ versus LN film thickness $h_{LN}$ and etching depth $h_{wg}$ at a fixed waveguide width $W_{wg}$ = 900 nm.

Fig. S10b compares $V_\pi$ trends for different $h_{LN}$ and $h_{wg}$. For a fixed $h_{LN}$, $V_\pi$ rises with $h_{wg}$: greater etching depth enhances lateral confinement but reduces the spatial overlap with the RF field, decreasing the voltage-induced index change. For a fixed $h_{wg}$, a thicker $h_{LN}$ yields a smaller $V_\pi$. However, when the ratio $h_{wg}$ / $h_{LN}$ is too small, the optical mode becomes predominantly confined in the slab layer (Fig. S9b), causing strong bending radiation losses (Fig. S11a). A sufficiently large $h_{wg}$ / $h_{LN}$ suppresses such losses (Fig. S11b and c). We therefore choose $h_{wg}$ = 200 nm and $h_{LN}$ = 500 nm (red star in Fig. S10b).

Two waveguide widths are used in the device. In the modulator section, a narrower width of 900 nm is adopted to minimize the electrode gap. For the bends, simulations (Fig. S11b and c) show that wider waveguides exhibit lower bending loss for a given film thickness; we therefore use a width of 2.3 μm for the bends. To avoid additional width transitions, the input/output waveguides of the multimode interference couplers (MMIs) and the delay waveguide are also designed with the same 2.3-μm width.

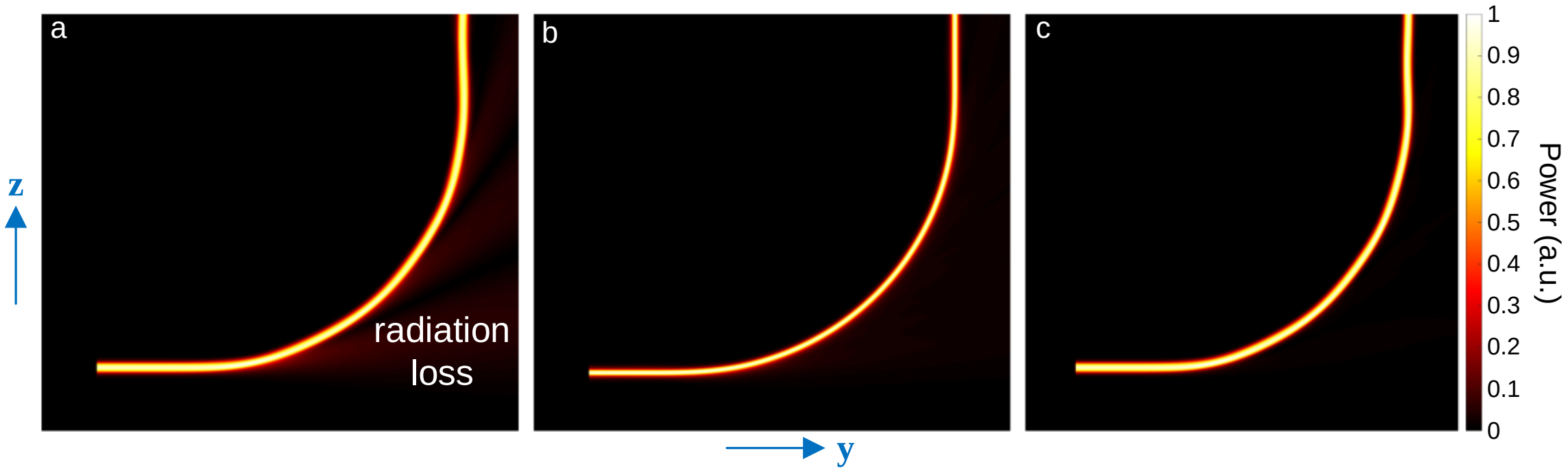

**Figure S11.** Bending transmission performance: (a) $h_{wg}$ = 200 nm, $h_{LN}$ = 600 nm, $W_{wg}$ =2.3 μm: radiation loss, transmission 92.7%; (b) $h_{wg}$ = 200 nm, $h_{LN}$ = 500 nm, $W_{wg}$ =900 nm: transmission 97.4%; (c) $h_{wg}$ = 200 nm, $h_{LN}$ = 500 nm, $W_{wg}$ =2.3 μm: transmission 99.7%.

The electrode structure is another key factor affecting modulation efficiency. In thin-film LN modulators, placing the metal electrodes as close as possible to the LN waveguide enhances the overlap between the RF electric field and the optical mode, thereby reducing $V_{\pi}$. However, if the electrodes are too close, the evanescent tail of the optical field may overlap with the metal, introducing additional metal absorption loss and increasing fabrication complexity and device instability. Positioning the electrodes on the upper cladding layer maintains a certain distance from the LN waveguide; this not only simplifies the fabrication process compared with direct deposition on the LN film but also allows a narrower electrode gap. Considering a typical lithographic misalignment of about 1 μm in fabrication, we choose a final electrode gap of 4 μm. This gap ensures low metal absorption loss even under misalignment, providing better process tolerance and transmission stability.

## Multimode interference coupler

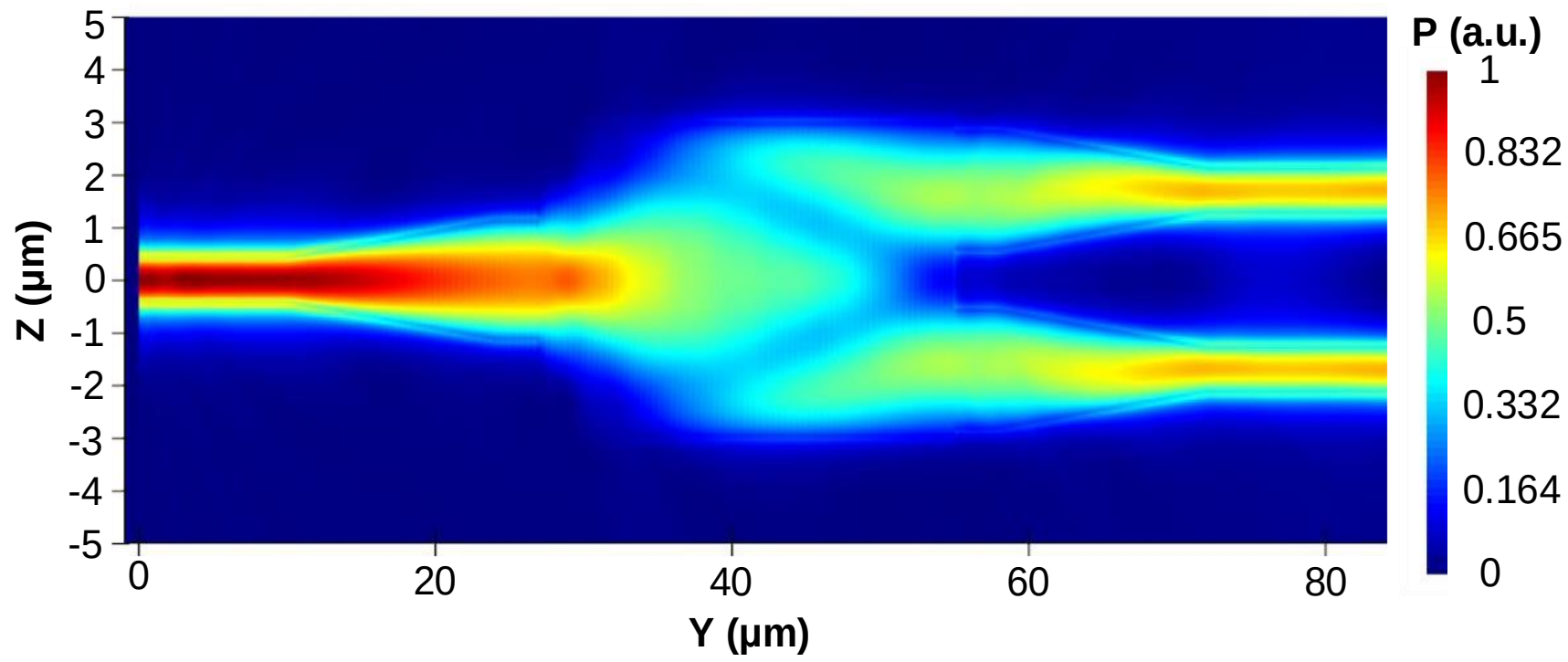

**Figure S12.** Simulation results of the 1×2 MMI.

We design the 1×2 MMI by first using eigenmode expansion (EME) solver for rapid parameter scanning, then verifying the optimized design with finite-difference time-domain (FDTD) simulation. The FDTD-calculated electric-field distribution (Fig. S12) yields a power transmission of approximately 0.49 from the left input to each of the two right output ports, essentially achieving the minimum insertion loss.

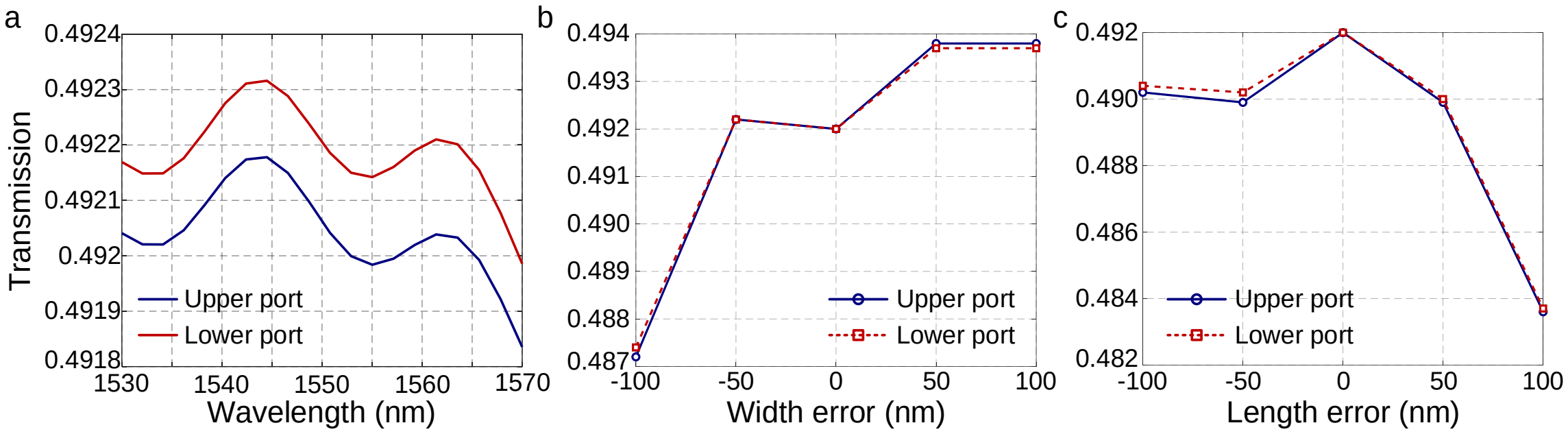

**Figure S13.** Tolerance analysis of the 1×2 MMI coupler to wavelength and fabrication errors. (a) Transmission of the two output ports versus wavelength; (b) Transmission versus MMI width deviation; (c) Transmission versus MMI length deviation.

Using an EME solver, we simulated the transmission of the MMI output ports under variations in wavelength, length, and width fabrication errors (Fig. S13). Across the 1530–1570 nm wavelength range, the transmission of both ports remains above 0.491. Within ±100 nm fabrication tolerances in both length and width, the transmission stays above 0.482 for both ports, with highly consistent behavior between the two ports under the same error conditions. These results demonstrate that our MMI design exhibits strong robustness to both wavelength variations and fabrication errors.

## Delay waveguide

The ordinary and extraordinary refractive indices ($n_o$ and $n_e$) of the LN film in the LNOI wafer were measured experimentally using a spectroscopic ellipsometer, and these measured values are used in all simulations. In Fig. 2a, the light propagates along the crystalline Z-axis in the delay waveguide, which has a width of 2.3 μm and an etch angle of 75°. The TE mode, polarized along the Y-axis, has a group index of $n_g = 2.326$. With a waveguide length of $L$= 67.5 μm, the corresponding delay at a center wavelength of 1557.7 nm is $T_D = L \times n_g / c$ = 523. 79 fs, where $c$ is the speed of light in vacuum. This length is chosen such that $T_D$ gives $N \approx 200$ in the relation $T_D = N \times T_c / 2 + t_d$, which, for $t_p$ = 1.2 ps, simultaneously provides large $TS_N$ and $LUI_T$ according to Fig. 1d and e.

## Edge coupler

To package the integrated DEST, suitable edge couplers are required to interface between the on-chip and fiber optics. Fig. S14 presents two mode-converter designs. In Fig. S14a, only the upper ridge section of the LN waveguide is tapered, whereas in Fig. S14b, the lower thin-film section is first tapered, followed by a second taper introduced in the upper ridge at a certain distance. Fig. S15 shows the simulated coupling efficiencies: approximately 43% for the single-layer taper and 60% for the double-layer design. The double-layer inverse taper is therefore adopted in the final fabricated device.

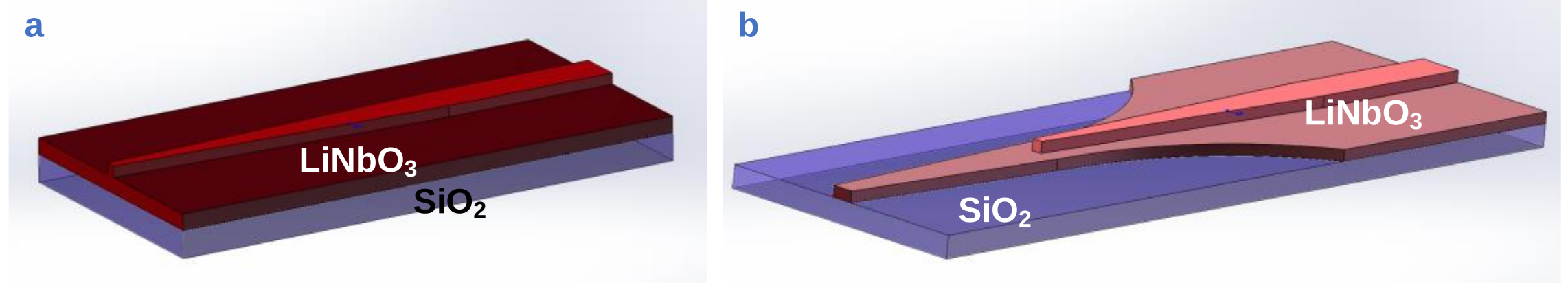

**Figure S14.** Design strategies for the edge mode converter: (a) single-layer inverse taper; (b) double-layer inverse taper.

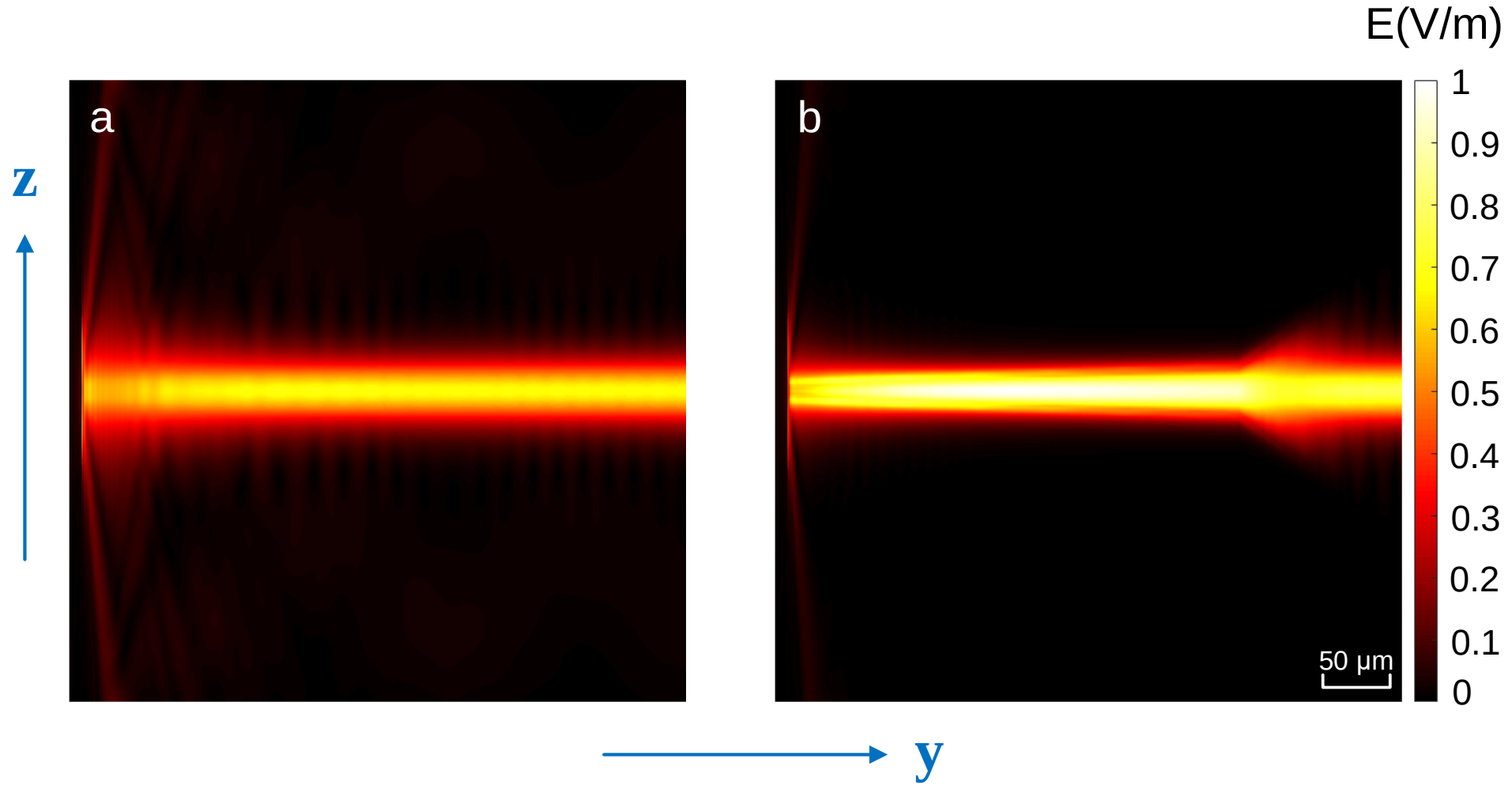

**Figure S15.** Simulation results of the edge mode converter: (a) single-layer inverse taper; (b) double-layer inverse taper.

## Half-wave voltage measurement

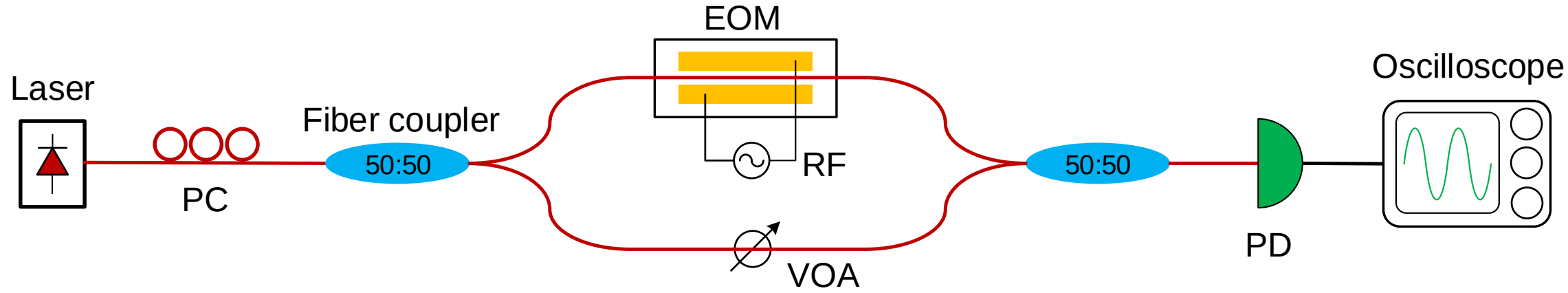

**Figure S16.** Setup for measuring the half-wave voltage of a single EOM. PC, polarization controller; VOA, variable optical attenuator.

The half-wave voltage of a single LN EOM is measured using an interferometric method, with the experimental setup shown in Fig. S16. Light from a laser is polarization-controlled by a polarization controller and then split into two paths by a 50:50 fiber coupler. One path passes through the EOM under test, which is driven by a periodic sawtooth signal; the other path serves as a reference arm with its power adjusted by a variable optical attenuator. The two paths are recombined in a second 50:50 coupler, and the resulting interference signal is detected by a photodetector and recorded by an oscilloscope. By scanning the drive voltage applied to the EOM and monitoring the corresponding changes in the interference pattern, the half-wave voltage can be extracted. The experimental measurement results are shown in Fig. 2g.

## III. Computational resource estimation

This section details the hardware resources and time overhead required to compute pairwise cross-correlations for a massively integrated DEST array. We first consider the case of a single chip with multiple channels, and then extend the analysis to large-scale arrays.

Assume an integrated DEST chip (Fig. 1g) provides 16 differential timing outputs, sampled by a 1 GSa/s ADC and fed into a field programmable gate array (FPGA). To achieve real-time cross-correlation and averaging of the 16 high-speed channels, the FPGA employs an efficient parallel frequency-domain processing architecture based on algebraic equivalence (Fig. S17).

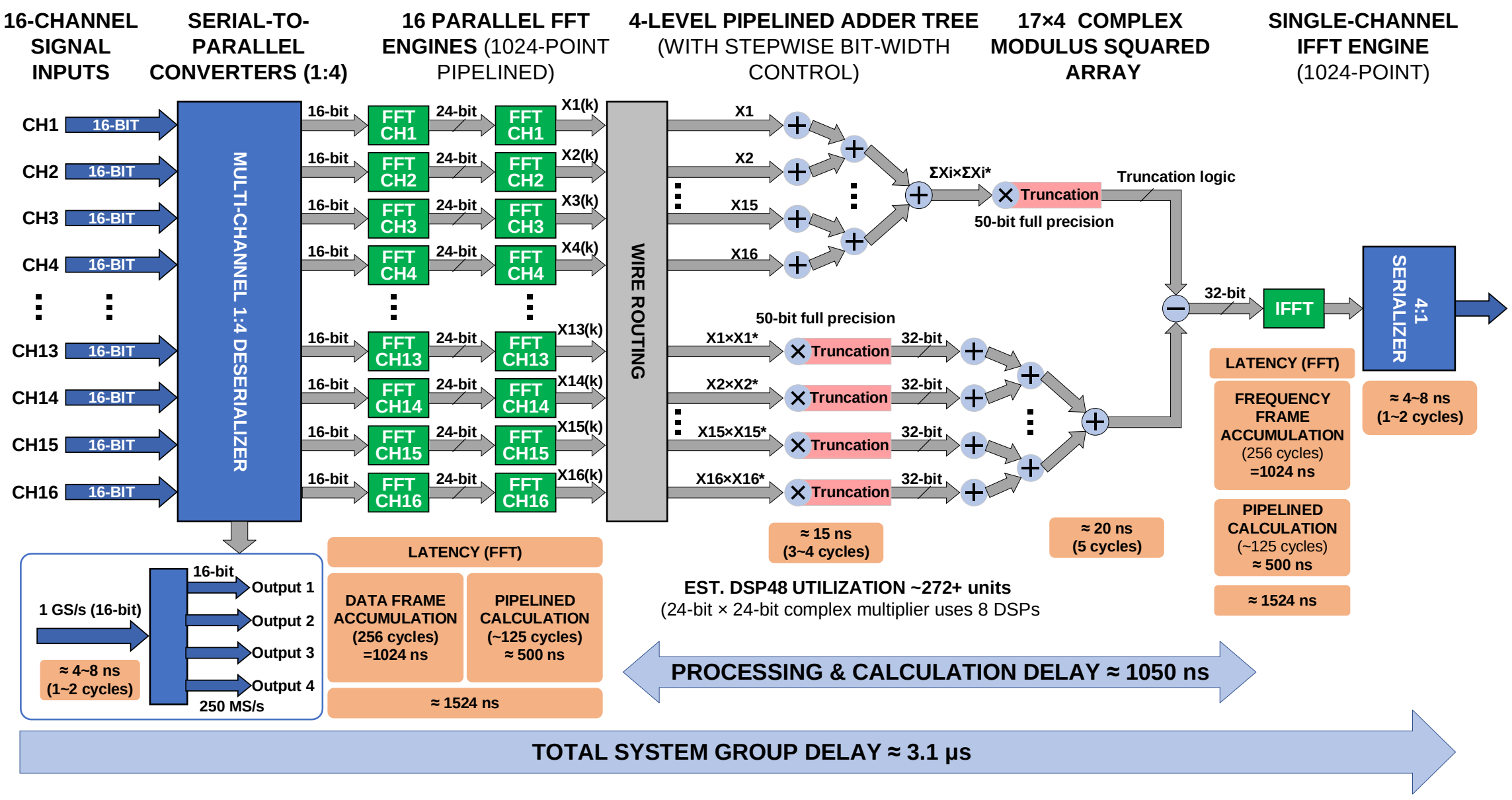

**Figure S17.** Flowchart of cross-correlation computation and delay estimation for 16-channel parallel input data on FPGA. Each channel after the multi-channel 1:4 deserializer is 4-way multiplexed; for clarity, only one multiplexed branch is illustrated. FFT, fast Fourier transform; IFFT, inverse fast Fourier transform.

Since the 1 GHz clock exceeds typical FPGA routing limits, each 1 GSa/s input stream is first deserialized by a 1:4 converter into four parallel 250 MSa/s sub streams. These parallel time-domain streams are synchronously fed into 16 dedicated pipelined 1024-point FFT engines. Each FFT engine adopts a 4-point parallel hardware architecture, processing four time-domain samples and outputting four frequency-domain samples per 250 MHz clock cycle. The FFT output bit width is extended to 24 bits to prevent weak signals from being buried by quantization noise.

In the frequency domain, the system uses Eq. (1) to split the data flow into two parallel branches. In the first branch, the 16 complex spectra are fused through a 4-layer complex addition tree into a single sum, from which the squared magnitude of the sum is computed. In the second parallel branch, each of the 16 spectra independently passes through a squaring unit to obtain its own squared magnitude, and these are accumulated via a real addition tree to yield the sum of squared magnitudes. The outputs of the two branches are subtracted, and the result is then truncated to obtain a high-precision 32-bit frequency-domain result. This result is sent to a single 1024-point IFFT engine, and finally a 4:1 serializer recombines the parallel outputs into a single 1 GSa/s continuous data stream for seamless real-time output.

The total group delay is approximately 3.1 μs, comprising unavoidable buffering latency (1024 ns for input frame accumulation and 1024 ns for output reconstruction, totaling 2048 ns) and pure computation delay (FFT and IFFT each ~500 ns, squaring and addition tree <40 ns, total ~1050 ns). Despite this fixed latency, the deeply pipelined architecture sustains full 1 GSa/s throughput without dead time.

The hardware implementation requires approximately 550 DSP units. This demand arises from several architectural decisions. The 24-bit FFT output requires cascading two DSP48E2 multipliers per 24-bit real multiplication. The squared magnitude of a complex number requires four real multipliers (2 for each of the two real squarings). With the 4-way parallel architecture, the squaring units in the two branches consume $1 \times 4 \times 4 = 16$ DSPs and $16 \times 4 \times 4 = 256$ DSPs, respectively, totaling 272 DSPs. Adding the 16 FFT engines and IFFT module brings the total to about 550 DSPs, a requirement readily met by mid-range FPGAs such as the AMD/Xilinx Virtex UltraScale+ VU3P.

If many DEST chips are integrated into an array (Fig. 1g), application specific integrated circuits (ASICs) can replace FPGAs, and with mass production the per-chip cost becomes negligible. To maintain 1 GSa/s throughput, the cross-correlation averaging can be computed via a multi-layer ASIC cascade. In this case, each DEST ASIC outputs only two frequency-domain sums: $\Sigma X_i$ and $\Sigma|X_i|^2$. Taking 12500 chips in 1 $m^2$ as an example (20 × 4 $mm^2$ footprint per chip), each group of 16 chips is summed through a four-layer addition tree, requiring 782 ASICs. Four cascade layers yield 782 + 49 + 4 + 1 =836 ASICs, with the final ASIC performing subtraction, IFFT, etc., resulting in a total latency of approximately $3.1 + 0.004 \times (4 + 4 + 4 + 2) = 3.156$ μs.

If the timing signal lies in the low-frequency band (e.g., gravitational wave), the final output need not maintain the 1 GSa/s rate. Multi-stage decimation can substantially reduce the data rate, enabling a few graphics processing units (GPUs) to compute the full pairwise cross-correlation average directly. Assuming a sampling rate of 1 MSa/s, the total data rate for $2\times10^5$ channels is 400 GB/s, which fits within a small server with eight 400 GbE network interface card (NICs) and eight GPUs. For a 1-s processing window ($10^6$ points per channel), the computational load is dominated by the FFTs. Assuming a 1024-point FFT, a real FFT of length $K$ requires approximately $2.5K\log_2(K)$ operations [Ref. S2]. The total FFT throughput is therefore $2 \times 10^5 \times 10^6 \times 2.5 \times \log_2(1024) = 5$ TFLOPS, well within the capability of a single consumer-grade RTX 4090. Thus, a modest server with eight such GPUs can easily handle the task.